\documentclass[preprint]{aastex}

\def\ltsim{ \,{}^<_\sim\, }
\def\gtsim{ \,{}^>_\sim\, }

\shorttitle{NGC 5128 Clusters}
\shortauthors{Harris et al.}

\begin{document}

\title{Structural Parameters for Globular Clusters in NGC 5128\altaffilmark{1}}

\author{William E.~Harris\altaffilmark{2}}
\affil{Department of Physics \& Astronomy, McMaster University,
    Hamilton ON L8S 4M1 }
\email{harris@physics.mcmaster.ca}

\author{Gretchen L.~H.~Harris\altaffilmark{2}}
\affil{Department of Physics, University of Waterloo, Waterloo ON N2L 3G1}
\email{glharris@astro.uwaterloo.ca}

\author{Stephen T.~Holland}
\affil{Department of Physics, University of Notre Dame, Notre Dame IN 46556-5670}
\email{sholland@nd.edu}

\and

\author{Dean E.~McLaughlin}
\affil{Space Telescope Science Institute, 3700 San Martin Drive, Baltimore MD 21218}
\email{deanm@stsci.edu}

\altaffiltext{1}{Based on
observations with the NASA/ESA {\sl Hubble Space Telescope}, obtained from
the Space Telescope Science Institute, which is operated by the Association
of Universities for Research in Astronomy (AURA), Inc., Under
NASA Contract NAS 5-26555.}

\altaffiltext{2}{Visiting Astronomer, Research School of Astronomy and 
Astrophysics, Australian National University, Weston ACT 2611, Australia}

\begin{abstract}

We present new imaging measurements of 27 individual globular clusters
in the halo of the nearby elliptical galaxy NGC 5128, obtained
with the {\it Hubble Space Telescope} STIS and WFPC2 cameras.  
We use the cluster light profiles
to determine their structural parameters (core and half-light
radii, central concentration, and ellipticity).  
Combining these with similar data for selected
inner-halo clusters from Holland et al.~1999 (AAp, 348, 418),
we now have a total sample of 43 NGC 5128 
globular clusters with measured structural properties.
We find that classic King-model profiles match the clusters extremely well, and
that their various structural parameters (core- and half-light radius,
central surface brightness, central concentration) fall in 
very much the same range as do the clusters in the Milky Way and M31.
We find half a dozen bright clusters which show tentative evidence
for ``extra-tidal light'' that extends beyond the nominal tidal radius, 
similar in nature to several such objects previously found in the Milky Way
and M31; these may represent clusters being tidally stripped, or possibly
ones in which anisotropic velocity distributions are important.  
We also confirm previous indications that 
NGC 5128 contains relatively more clusters with large ($\epsilon > 0.2$) ellipticity
than does the Milky Way.  Instead, the $\epsilon-$distribution
of the NGC 5128 clusters strongly resembles that of the old clusters in the LMC
and also in M31.
Finally, calculations of the cluster binding energies $E_b$ as defined by  
McLaughlin 2000 (ApJ, 539,618) 
show that the NGC 5128 clusters occupy the same 
extremely narrow region of the parametric ``fundamental plane'' as do their
Milky Way counterparts.
Our data are thus strongly consistent with the claim that the
globular clusters in both NGC 5128 and the Milky Way
are fundamentally the same type of object:  old star clusters with similar
mass-to-light ratios and King-model structures.

\end{abstract}


\section{Introduction}

Globular star clusters have remarkably simple structures that are
well approximated by isotropic, single-mass \citet{kin66} models.
In the multi-dimensional space of all
their structural quantities such as scale radii, central concentration,
surface brightness, velocity dispersion, mass-to-light ratio, and so
forth, it is striking that real globular clusters in the Milky Way
inhabit only a narrow region referred to as 
the fundamental plane \citep[FP; see][]{djo95}.  

Recently, \citet{mcl00} has shown that a particularly simple way of expressing
the FP is to note that any King model is completely specified by
four input parameters such as
total cluster luminosity $L$, central concentration
$c =$ log $(r_t/r_c)$, mass-to-light ratio, and binding energy
$E_b$.  Adding in the two strong empirical constraints that $M/L \simeq$
const and $E_b \sim L^2$ then requires the clusters to lie on a
two-dimensional slice of this 4-space, leaving only two quantities
($c$ and $L$) to determine the residual scatter on this FP.
In turn, the concentration $c$ is correlated with $L$, leaving
the remarkable result that the structures of these clusters are
fixed largely by just one major {\sl internal} parameter, their
total mass (or luminosity, at a given age).  The {\sl external} environment,
i.e.~its location in the Galactic tidal field, also has some
influence on the FP parameters.

However, clusters may have formed under drastically different environmental 
conditions in other galactic environments 
(giant and dwarf ellipticals, starburst systems, galactic bulges and 
rings, etc.).  The structures of clusters in these other types of galaxies
remain to be investigated. 
The only other large galaxy for which this kind of study has been carried out 
in detail is M31, a large disk galaxy much like the Milky Way, and perhaps not
surprisingly, its globular clusters strongly resemble those of the Milky Way 
\citep{fus94,hol97,bar02}.

Beyond the Local Group, the nearest large galaxy containing many 
globular clusters available for detailed measurement
is NGC 5128, the giant elliptical at the center of the Centaurus group
at a distance $d \sim 4$ Mpc.
More importantly, it is a very different type of galaxy than any
in the Local Group, and quantitative study of its clusters holds
considerable promise for adding constraints to formation
modelling.  In addition, a point of special interest from the viewpoint of globular
cluster structural studies is that, because of sheer population size,
NGC 5128 has many clusters at the
upper end of the globular cluster mass distribution ($\gtsim 10^6 M_{\odot}$);
with a total population of perhaps 1900 clusters \citep{har84},
it should have 40 to 50 clusters with $M_V \ltsim -10$.
By contrast, all the Local Group galaxies combined contain perhaps $\sim 600$
globular clusters and thus have only a few with $M_V \ltsim -10$.
NGC 5128 thus gives us the opportunity, in a single galaxy, to explore the empirical FP 
relations at extremely high cluster mass
approaching $10^7 M_{\odot}$ ($M_V \sim -12$).

Clusters in galaxies as distant as Centaurus 
appear barely nonstellar under typical
ground-based imaging resolutions of $1''$ but can be much more
well resolved with the cameras on HST.  The first clear demonstration
that accurate structural profiles and King-model 
parameters could be obtained for these
objects was provided by \citet{har98}, who studied a single
outer-halo cluster in NGC 5128.  Shortly afterward, \citet{hol99}
obtained similar results for a selection of inner-halo clusters.
In this paper, we present new imaging data for another sample
of clusters in this important galaxy, more than doubling the
total sample.

Throughout this paper we assume $d = 4.0$ Mpc for NGC 5128
and a foreground (Milky Way) reddening of $E_{B-V} = 0.11$, for an apparent
distance modulus $(m-M)_V = 28.35$ \citep{har00}.  At this distance,
1 arcsecond is equivalent to a linear scale of 19.4 pc.
Our adopted distance is a mean of the results from three methods
including the red-giant-branch tip luminosity \citep{har99}, the planetary
nebula luminosity function \citep{hui93}, and the $I-$band surface brightness
fluctuation technique \citep{ton01}.  The mean distance is likely to be uncertain to
$\pm0.2$ Mpc based on the close mutual agreement of these methods.

\section{Observational Material and Sample Definition}

During the course of an HST Cycle 9 SNAPSHOT program (GO-8664) we obtained
new images centered on 18 known globular clusters spread throughout 
the halo of NGC 5128.
These were all 400-second exposures with the STIS camera 
in its unfiltered 50CCD/CL imaging mode and with gain=1 $e^-$ per du.
In addition to the STIS material, we also had in hand 
long-exposure WFPC2 $(V,I)$ images (gain = 7) of four clusters
from our previously published photometric studies of the halo red-giant stars
in this galaxy \citep{har98,har02}.
Two objects, G302 and the newly discovered cluster C100 (see below), 
were imaged with both STIS and WFPC2, providing a small
but useful consistency test of our structural fitting routines.

The candidate objects were all selected from the photometric list of known 
NGC 5128 clusters analyzed by \citet{har92}; our identification
numbers (C and G prefixes) follow their list and are given
in Table \ref{coords}.  All but one of our clusters 
were imaged either at the STIS scale of $0\farcs0507$/px or the
PC1 scale of $0\farcs0455$/px; the one remaining object (G19) fell 
on the lower-resolution $0\farcs0996$/px scale of the WF3 camera.

All of the individual STIS and WFPC2 frames were inspected carefully to locate
any clearly nonstellar objects in the field regardless of brightness, 
in addition to the previously known clusters.
Almost two dozen additional cluster candidates not in any
previous lists were located this way.
On closer inspection, most of the ``extra'' candidates proved to be 
more likely to be faint background galaxies, with light profiles
that were asymmetric, highly extended, or lumpy. 
However, seven were found to have characteristic sizes and symmetric
King-type profiles like those of globular clusters and are thus
kept in our sample. We believe these to be newly discovered
clusters in NGC 5128.  Our final list of 27 objects which we regard as
definite or highly probable globular clusters in NGC 5128 is presented
in Table \ref{coords}; the seven new objects are assigned
ID numbers C100 - C106.

In Figure \ref{samples}, we show portions of the STIS images to illustrate
the appearances of three of our measured clusters having a range of 
luminosities and galactocentric distances.  As will be shown more
quantitatively below, the cluster images are well resolved compared
with the profiles of stars (PSFs).

\section{Photometry}

The first step in our analysis was to take the preprocessed images
supplied by the STScI archive and derive positions and total magnitudes
for each object.  The coordinates were obtained directly from the
information in the image headers and the STSDAS {\sl xy2rd} routine
and are expected to be accurate to $\pm1''$.  These are listed in
columns (2) and (3) of Table \ref{coords}, along with their projected
galactocentric distance $R_{gc}$ (column 3) under the assumption $d$(NGC 5128) = 4 Mpc.
The final column of the Table gives the particular detector (STIS, PC1, or WF3)
on which the cluster was imaged.

For measurement of total magnitudes, we 
constructed curves of growth (integrated magnitude vs. aperture radius)
from concentric-aperture photometry to select
an aperture radius large enough to enclose almost all the
cluster light without being dominated by sky noise.
We adopted  $r = 60$ px $\simeq 3''$ (as will be seen below,
at this radius we reach the outermost extreme at which we can trace the
surface brightness profile for most of the clusters).
For a few cases where the objects were crowded by neighboring
stars, we used the largest feasible uncrowded aperture radius
and then applied the fiducial curves of growth to make the (small) necessary
extrapolations out to $r = 3''$.

For the four clusters (C44, G19, G302, C100) on the WFPC2 frames, converting
the integrated instrumental magnitudes into $V$ and $I$ was a straightforward matter
of applying the usual calibration equations \citep{hol95} for the relevant
filters (F606W, F814W).  Their resulting $(V,V-I)$ magnitudes and colors
are listed in columns (5) and (6) of Table \ref{coords}.

For the much larger set of objects measured on STIS, 
the unfiltered (``clear'' or CL) 50CCD configuration of the camera
has a central wavelength near that of the standard $V$ magnitude
and thus can be transformed moderately well into $V$.
Here we label the 50CCD ``instrumental'' magnitude as
$v_{CL} \equiv -2.5$ log $f_{CL}$ 
(where $f_{CL}$ in DN/sec is the measured flux from the object).
However, the zeropoints $(V-v_{CL})$ quoted in the literature
\citep[e.g.][]{rej00,gar00},
as well as the SYNPHOT method using PHOTFLAM from the image header, 
differ by 0.2 mag or more from one source to another.  To set the 
STIS photometric zeropoint for this particular study
we rely instead on two {\sl local} calibration
methods which employ the clusters themselves.
Since these calibrating objects are exactly the same entities
as our program objects, this approach also minimizes any systematic
error due to nonzero color terms in the transformation.
First, we used the small set of $(V,V-I)$ measurements for clusters
obtained by \citet{ton90}; comparing our list 
with theirs revealed 11 clusters in common.
Second, we used the Washington photometry $(C,M,T_1)$ of the clusters
published by \citet{har92} along with the established conversion into $V$
\citep[see][]{gei96},
\begin{equation}
V \, = \, T_1 \, +\, 0.052 \, + \, 0.256 (C-T_1) \, \, .
\end{equation}
This approach gave $V$ magnitudes for 16 clusters in our observed list.

In Figure \ref{Vcalib} we plot the difference $(V - v_{CL})$
against $V$.  The Tonry/Schechter sample gives
$\langle V-v \rangle = 26.34 \pm 0.02$ with an rms dispersion $\pm 0.075$ mag,
while the transformed Washington photometry gives 
$\langle V-v \rangle = 26.27 \pm 0.02$ with
dispersion $\pm 0.060$ mag.  Though there is a 
noticeable difference between the two methods, we
have no basis to prefer one strongly over the other.\footnote{From a more
extensive comparison of the converted Washington photometry with the 
Tonry/Schechter sample, \citet{har92} found a mean 
offset $\langle V_{TS} - V_W \rangle \simeq 
0.04$ mag, very similar to our net difference of $0.07 \pm 0.03$.}
Averaging all the data together, we adopt
$\langle V-v \rangle = 26.29 \pm 0.05$.  In the end,
it should be recognized that the STIS magnitudes cannot
be as accurately calibrated as normal ones that are filter-defined,
but they are expected to be useful to approximately the accuracy quoted.
It is encouraging that, for our only uncrowded object measured with
both WFPC2 and STIS (cluster G302), the calculated STIS/CL magnitude agrees
with the WFPC2 one to 0.05 mag.  The agreement is not as close for
cluster C100, also measured with both STIS and PC1, but 
this object is severely crowded by a star of comparable 
brightness only $1''$ away.

\section{Structural Parameters and Model Fitting}

The model fitting was carried out in two partially independent ways
as follows:

\noindent (1) We performed ``two-dimensional''
fits whereby a \citet{kin66} model with a given 
set of parameters $(W_0, r_c, c, \epsilon)$ (central potential
$W_0$, core radius $r_c$, central concentration $c = $ log$(r_t/r_c)$,
and ellipticity $\epsilon = 1 - b/a$) is calculated, and then convolved with the full 
two-dimensional PSF shape for the given image.
The PSF profile was determined empirically from several bright
uncrowded stars on the frames.
The seeing-convolved model is
then matched to the surface brightness of the
raw cluster image after subtraction of the
background light, and the King model parameters are varied to
achieve the best match to the data. 
The procedure is described fully in \citet{hol99}.

\noindent (2) For comparison, we used the empirical ellipse-fitting code
in {\sl STSDAS} ({\sl analysis.isophote.ellipse}) to generate
smoothed profiles for all the candidate objects along with 
estimates of the surrounding local background light intensity.
The {\sl ellipse} code expects only that the light profile of the object
is ellipsoidal and decreasing monotonically outward.
In carrying out the fits we allowed the ellipticity
and position angle of each annulus to vary but kept the ellipse center fixed.
By subtracting the model from the original image and inspecting the
residual light, we found that the background surface brightness $b_V$
could be determined (on the STIS images) with a typical uncertainty
of $e(b_V) \simeq 3$ DN (or $\pm 3 e^-$ for a gain of 1 $e^-$/DN);
changes in the background 
larger than that above or below the nominal $b_V$ level
left clearly distinct outer ``edges'' 
once the profile was subtracted from the cluster.
In this way, we found that we could trace the cluster light profiles
down to a level within about 3\% - 5\% of the sky background.
Attempts to probe still fainter proved to push the measurements
on these relatively short-exposure frames
past their reliable limits.
If higher S/N data are obtained in future, however, it should be possible to
study the outermost envelopes of these clusters more thoroughly.

The {\sl ellipse} code generated one-dimensional 
lists of surface brightness vs. semi-major axis.  These
were then matched to 1-D King model profiles 
convolved with the PSF profile.
Sample profiles and best-fitting King models are shown 
in Figure \ref{profiles} for four clusters covering a wide range of brightness.
The seeing-convolved model profiles
match the clusters quite well particularly in the inner and middle
radial range where the cluster light is much brighter than the background.
The only exceptions were a few of the very faintest objects for which
the signal-to-noise was lowest and the contrast over background the smallest.
Compared with the stellar PSF, which has a FWHM of $0\farcs11$,
the cluster profiles are typically about three times broader
and thus very well resolved.

Sample illustrations of the quality of fit of the generic King models
are given in Figure \ref{resid}, where (for the same four clusters as
in Figure \ref{profiles}) we show the residuals after subtraction of the
model profile from the {\sl ellipse} code measurements.  At an intensity
level about one magnitude fainter than the background level, the internal
uncertainties of the measured points become very large and do little
to constrain the fit.  At smaller radii, however, the model adheres closely
to the data, with typical scatters of $\ltsim 0.05$ mag per data point.
For comparison (see below), on these graphs the core radii of the clusters 
would fall typically at log $r_c \sim -1.3$ and the half-light radii
at log $r_h \sim -0.4$, well within the higher-precision part of the fit.

In general we found that the two methods outlined above agreed well,
though the 1-D fits tended to be numerically more robust (in several
cases, particularly for faint objects, the 2-D fit did not converge
satisfactorily).  
The adopted mean values of the structural parameters
for all of our program objects, from the average of the two methods,
are summarized in Table \ref{parameters}.  In the Table, we also include the 
half-light radius $r_{h}$ as calculated from the model fit, since it is of interest
as a quantity which is relatively immune to dynamical evolution
within the cluster.  The {\sl projected} half-light radius $R_h$, a more
conventionally used observational quantity, can be obtained
from the (three-dimensional) quantity $r_h$ by the numerical approximation
$R_h \simeq 0.73 r_h$, which is accurate to a few percent over the range
of $c-$values that apply to normal clusters.

The mean uncertainties
in each measured quantity are listed at the bottom of the Table.
\citet{hol99}, through extensive numerical tests including
simulated clusters, found that the expected measurement uncertainties
for all quantities were 5\% - 10\% depending on brightness.

For the two clusters G302 and C100 (the latter object being
crowded by a bright star, as noted above) measured on both the STIS and WFPC2 data,
the independently determined model fits gave structural parameters
in good mutual agreement (see Table \ref{parameters}).  For these two, we adopted
the core radii and ellipticities from the WFPC2 images since they were much longer
exposures and thus higher S/N.

The central surface brightnesses $\mu_V^0$ ($V$ mag per arcsec$^2$) listed
in Table 2 are partially ``indirect'' measurements in the sense that they are
the surface brightnesses of our best-fit King-model curves read off
at zero radius.  We stress that the innermost core structures at $r << r_c$ cannot
be truly resolved for any of our program objects.  Thus (for example)
if any of them is in fact a core-collapsed object with a power-law
profile for the inner core, we would significantly underestimate $\mu_V^0$
and overestimate $r_c$.

The last column of the Table gives the position angle $\theta$ of the
isophotal major axis, measured counterclockwise (eastward) from North.
The best-fitting orientation angle often varied by $\pm 20^o$ or so
from the inner parts of the cluster to the outskirts, but our quoted mean
refers to the position angle near the half-light radius $r_h$ (see below).
These orientation angles are more uncertain for clusters of smaller
ellipticity.  We find no statistical preference for any particular cluster orientation.

\section{Clusters With Extratidal Light?}

In the course of examining the profiles of all the individual
clusters, we noted that for some of the brighter clusters
in the sample, the fitted King models fell clearly below the 
measured cluster profile at the largest radii, 
even though the model fit was extremely accurate
at smaller radii.  In other words, within the context of the
formal model fits (which are based on the assumptions of isotropic
velocity dispersion and a single-mass stellar population) these
clusters exhibit ``extratidal light'' (XTL) continuing
outward past the nominal tidal radius $r_t$.  Illustrations of this
effect for four objects are shown in Figure \ref{tidaltails}.  
Although any individual data point at large radius is quite uncertain
(see Fig.~\ref{resid}), these half-dozen clusters are among the brightest
and have the best-determined profiles of the entire sample.

To test the reality of this phenomenon, we went through the numerical
exercise of arbitrarily adjusting the adopted background intensity
level $b$ until the ``extra'' light at large
radius essentially vanished and the residual outer profile approximately fit the
King model at all radii.  The size of this arbitrary shift $\Delta b$ was
then compared with the true external uncertainty in the background
(typically $e_b \simeq \pm 3$ DN, as noted above).  For most
clusters in the sample, the necessary shift $\Delta b$ was $\ltsim 2 e_b$
and thus the significance of the effect was marginal at best;
that is, we found that {\sl small} positive or negative
deviations relative to the King models at large radii
could be readily understood as slightly incorrect adopted backgrounds.
However, for six clusters (listed in Table 3), we found $\Delta b > 3 e_b$
and for these we cautiously suggest the presence of a real XTL component.

All six of these clusters are brighter than $M_V = -10$ and thus
are comparable with $\omega$ Centauri and NGC 6715, the two
most luminous clusters in the Milky Way.
For these six objects we subtracted the King model from
the observed profile to derive the amount of residual light present
in the XTL ``tail''.  In Table \ref{xtl.tab}, we list the projected
galactocentric distance $R_{gc}$; the cluster luminosity
$M_V$ (including the XTL); the increase in background light
$\Delta b$ necessary to artificially remove the anomaly; 
and the fraction of the 
total cluster light which is in the extratidal component.  
Although they are only marginally significant, these fractions 
range from 7\% to 17\% of the cluster luminosity and thus represent
a significant part of their entire stellar population.
The candidate XTL objects show no preference for any particular halo
location, with galactocentric distances ranging from 6 to 22 kpc.

The shapes of these extratidal profiles strongly resemble what has
been found for a few clusters in both the Milky Way \citep{gri95,leo00}
and M31 \citep{gri96}.  If the XTL is real, it may then
represent tidally stripped or evaporated stars now drifting away from
the cluster.  Deeper imaging data than we have
at present may be able to trace these faint components further outward
along thin streamers which would mark out the clusters' orbital paths.

Another interpretation discussed in the recent literature
\citep[e.g.][]{lee99,maj00,hug00,car00,hil00,bek01,mey01} 
is that objects like these may be
the luminous, compact nuclei of former dwarf satellite galaxies
(dE,N systems) that were accreted long ago.  In this case, the XTL
would be interpreted as the residual trace of the dE field-star population.
This idea has gained additional impetus from the observation that
NGC 6715 (the second most luminous cluster in the Milky Way halo)
is near the center of the disrupting Sagittarius dwarf \citep[e.g.][]{bas95,dac95,lay00}
and also from a variety of recent observations of $\omega$ Cen
indicating multi-epoch star formation (cf.\ the references cited above).

It is probably not surprising that several clusters in our
sample (6 clusters out of 27 observed) show these extended envelopes,
because our entire sample of clusters is biased in favor of such objects.
As noted above, our candidates for STIS imaging were
drawn from a list of known clusters \citep{har92} which were in turn
identified from ground-based photographic plates on the basis
of nonstellar appearance.  Thus the objects in our list should
be expected to be biased towards {\sl bright clusters with extended
envelopes} which were easiest to pick out as nonstellar under
$\simeq 1'' - 2''$ seeing.  (One of the clusters we find to have an
extended envelope, number C7, is
the first cluster identified in NGC 5128 by Graham \& Phillips 1980).

An alternative and more conservative interpretation of these extended profiles
is that the outer parts of these clusters actually do not consist of
``excess'' light, but might be fitted instead by models incorporating
multi-mass stellar populations and/or an anisotropic velocity distribution.
Either or both of these steps would allow fitting of a greater range
of projected density profiles \citep[e.g.][]{gun79}.  
Higher-quality data than we have at present will be needed to discriminate
more clearly among these alternatives.

\section{Structural Parameters and Correlations}

The globular clusters we have observed in this program are scattered
throughout the NGC 5128 halo, with projected galactocentric distances
ranging from $\sim 6$ to 25 kpc.  \citet{hol99} have obtained structural
parameters for 21 more objects projected on the {\sl inner} halo of
the galaxy ($R_{gc} < 3$ kpc), so by combining their sample with ours we can trace out
any trends in the structural parameters over a much wider $R_{gc}$ range.

From the Holland et al.\ list, we more or less arbitrarily
reject five objects with extreme colors
(4 with $V-I > 1.8$ which are likely to be either very heavily reddened
globulars or background galaxies, and one very blue object with $V-I = 0.4$).
All of these five have colors quite different from the $(V-I)\sim 1.0 \pm 0.2$
level characterizing all known globular clusters.
The remaining 16, added to our 27, give a total list of 43 definite or
highly probable globular clusters with measured King parameters.
The major noteworthy difference between the two sets of objects is
that the 16 from \citet{hol99} were all measured with rather short
WFPC2 exposures (mostly 180 sec), with higher background light,
and on the low-resolution WF2, 3 or 4 detectors.
Thus they are taken from lower-S/N data and with poorer image sampling
than were the majority of the objects in our new program.

\subsection{Scale Radii}

We first briefly compare our NGC 5128 clusters with those in the Milky Way,
using data from the 1999 edition of the \citet{har96} catalog.
As did \citet{hol99}, we restrict the Milky Way sample to 78 clusters
more luminous than $M_V \simeq -6.5$, with
reddenings lower  than $E_{B-V} = 1$,
and further from the Galactic center than 1 kpc, to make it as similar
as possible to our NGC 5128 sample.  Histogram comparisons of core
radius and ellipticity are shown in Figures \ref{rc_comp} and \ref{e_comp}.

The core radii for clusters in both galaxies fall very much in
the same global range ($r_c \ltsim 4$ pc for almost all objects),
and the medians of the two distributions are also similar
(median $r_c = 1.01$ pc for NGC 5128, 1.16 pc for the Milky Way).
However, the Milky Way sample has relatively more objects at very small
$r_c$.  This difference is likely to be due to a combination of selection
effects:  as noted above, many of the NGC 5128 candidate clusters were
pre-selected for their nonstellar appearance on ground-based images and thus
the sample is already biased towards extended structure.
In addition, it becomes increasingly
difficult on numerical grounds to measure the true value of a core radius
less than $0\farcs025$ (0.5 pc) given that it is convolved with a PSF
profile with HWHM $ \simeq 0\farcs05$.  We suggest that there are likely
to be many highly compact clusters with small core radii in NGC 5128 still
waiting to be found.

Our NGC 5128 sample also lacks objects with {\sl large} core radii $r_c \gtsim
5$ pc, a few of which exist in the Milky Way comparison diagram.  Although
the current statistical sample is small, we believe that this discrepancy 
might well be due to selection effects as well.  Most of
these Milky Way objects are moderately faint, diffuse objects in the outer
halo, and it is entirely likely that such objects would not have been noticed
in the photographic image surveys from which our target list was selected
(see above).  

\subsection{Ellipticities}

Our cluster ellipticity measurements are shown in Figure \ref{e_comp}
($\epsilon = 1-b/a$ where $b/a$ is the minor-to-major axial ratio).
These are weighted averages by annular luminosity over the measured radial range of the
model fits and correspond roughly to the $\epsilon-$values at the ``half-light'' 
radius $r_h$ commonly used in the literature.  Often the fitted ellipticity
is found to be a weak function of radius, though none of the clusters exhibited
dramatic ellipticity gradients.
In the second-last column of Table 2, we indicate whether the cluster showed
a clear increase in $\epsilon$ with $r$ ($+$), a decrease ($-$), or no
significant gradient (blank column).

Returning to Figure \ref{e_comp}, we note that
NGC 5128 has a rather flat, or modestly declining, $\epsilon-$distribution over
the range $0 < \epsilon < 0.3$.\footnote{The
lack of NGC 5128 clusters at {\sl very} small ellipticity $e < 0.04$ is likely
to be an artifact of incomplete removal of the slightly elliptical PSF,
as also noted by \citet{hol99}.  For larger $\epsilon$, it is less clear
whether or not any particular subtle
biasses or selection effects may exist in our sample.
The clusters were identified as members of NGC 5128 either on the basis of
radial velocity, or on the basis of their nonstellar appearance from
lower-resolution ground-based images in which their slightly out-of-round
shapes would have been at least partly washed out.  However, a much more 
complete candidate list will be needed before these issues can be addressed
more securely.}
From their small inner-halo sample of clusters, \citet{hol99} suggested 
that NGC 5128 had a much higher fraction
of high-ellipticity clusters than did the Milky Way \citep{whi87}.
We find that the medians for the two samples
are at $\epsilon = 0.11$ for NGC 5128 and 0.05 for the Milky Way, 
and a standard K-S test shows
that the two distributions differ significantly at higher than 99\% confidence level.
In short, we draw the same formal conclusion here as did Holland et al.  However,
we do not place excessive weight on it at this stage because of lingering
uncertainties about possible selection effects in our samples, which still 
represent a very small fraction of the entire NGC 5128 cluster population
(see the preceding footnote).  In addition, the ellipticities for the Milky Way
clusters follow from a somewhat different measurement process \citep[see][]{whi87}
and the possibility of small systematic differences remains a concern.

The observed ellipticity must be a complex product of the initial structure
of the cluster and its subsequent dynamical evolution but, as \citet{mey97}
comment, ``our understanding of this problem is rather patchy''.  
Flattening is almost certainly caused primarily by global rotation
of the cluster rather than internal properties such as velocity
anisotropy or external ones such as tidal elongation \citep{lag96,mey97,bar02}.  
Recent dynamical models indicate that internal relaxation coupled to the
external tidal field will in most
cases drive a cluster towards rounder shape 
over several relaxation times $t_{rh}$ \citep{fal85,lag96,lon96,ein99}.
Evolution is generally faster at higher rotation, lower central potential $W_0$, or
smaller galactocentric distance $R_{gc}$.  Notable, but not
overwhelming, evidence that this type of evolution actually occurs for real clusters 
may be seen in the LMC clusters, also shown in Fig.~\ref{e_comp}.
The shaded part of the LMC
histogram shows the $\epsilon-$distribution for 17 ``old'' clusters for which the
evolutionary states should be more closely comparable with the Milky Way
globulars (specifically, these 17 are in SWB age classes
VI or VII). The unshaded part
is for 48 younger clusters \citep[data from][]{kon89,fre82,gei80}.  
The median ellipticities are
$0.12 \pm 0.02$ for the older clusters and $0.16 \pm 0.01$
for the younger ones \citep[see also][who argue for a similar trend]{fre82}.  

The $\epsilon-$distribution for the LMC clusters has
long been realized to be different from the Milky Way \citep[][among 
others]{han94,kon89,kon90,bha89,van84,fre82,gei80}. Although an obvious problem
with making such a comparison is that
the great majority of the Magellanic objects are ``young'' ($T \ltsim 5$ Gyr) 
relative to the globulars in M31 and the Milky Way and thus less dynamically
evolved, it appears to be true that 
Magellanic clusters of all ages have $\epsilon$ predominantly $ \gtsim 0.1$.  
Interestingly, the old-LMC group statistically
resembles our NGC 5128 sample (they are different at less than 
70\% significance from a K-S test) much more closely than does
the {\sl younger}
LMC group (which differs from NGC 5128 at more than 99\% significance).

Lastly, in Figure \ref{e_comp} we also show comparisons with clusters in M31,
where the data are drawn 
from the high-resolution HST-based imaging study of \citet{bar02}.
For the M31 sample, we exclude 
four very faint objects with $\epsilon > 0.3$ to keep the same luminosity range
$M_V \ltsim -6$ that we use for the Milky Way and NGC 5128.  Even without these
few high-ellipticity clusters, however, 
the NGC 5128 and M31 samples are also remarkably similar
(the difference has less than 50\% significance from a K-S test).

The M31/LMC/N5128 comparison casts some doubt on the hypotheses that
either age or the strength of the external tidal field are the only,
or major, factors determining the $\epsilon-$distributions:
instead, the data in these three quite different galaxies
show that it is not unusual for very old clusters
to have preserved clearly elliptical shapes ($\epsilon \gtsim 0.2$) over
a Hubble time and regardless of their present environment.  
Yet another recently found example of a luminous globular cluster
with high ellipticity is object 13 in the lenticular galaxy NGC 1023,
with $\epsilon \simeq 0.37$ \citep{lar01}.
Of the four galaxies shown here, the Milky Way is 
the one which stands out as ``different'' from the other three.
More comprehensive datasets (more galaxies, and more clusters per galaxy)
will be needed to gain a better idea of the significance of this
preliminary result.

In a previous paper \citep{hh01}, we have noted
that the metallicity distributions for the halo stars in NGC 5128, M31, and
the LMC are also remarkably similar.  The $\epsilon-$distributions
may constitute additional evidence for our speculation that 
much of the stellar content
of NGC 5128, and M31, formed within Magellanic-sized pregalactic units before
assembling into the giant galaxies we see today.

\subsection{Correlations with $L$ and $R_{gc}$}

In Figure \ref{mu_v0}, we show the central surface brightness of our
clusters from Table 2, plotted against core radius $r_c$ and central
concentration $c$.  Here, the $\mu_V^0$ values have been corrected for
foreground absorption, which we adopt as $A_V = 3.1 E_{B-V} = 0.34$.  
Since our NGC 5128 sample is restricted mostly
to the brighter ones in the galaxy, we are probing only a small part
of the whole range of this correlation.  Nevertheless,
just as for the other cluster properties 
{\sl except} possibly for the ellipticities (see above), the central 
surface brightnesses fall in the same broad regions of these
scatter plots classically defined by the
more well studied globular clusters in M31 and the Milky Way (shown in
Figure \ref{mu_v0} by the dashed lines, adapted from Figure 6 of
Barmby et al.~2002).  The scatter in the observed distribution arises
from the spread in cluster luminosity $L$ at a given $r_c$ or $\mu^0$
(see Figure 19 of McLaughlin 2000a and his accompanying discussion).
The only notable exceptions are the two faintest
objects in our set, C102 and C105, which have anomalously high central concentrations.
However, the $c-$values for these faint objects
are very uncertain, and the possibility remains that these two may 
be background galaxies which crept through our selection
criteria.  Higher-S/N images and multicolor
measurements will be needed to confirm their identities.

In Figure \ref{mv_corr} we show the scatter plots of
core radius $r_c$, half-light radius $r_{h}$, central concentration index
$c$, and ellipticity $\epsilon$ with cluster luminosity.
The solid lines in each panel show the least-squares fits, which have the
following slopes:
$\Delta$log $r_c / \Delta M_V = 0.02 \pm 0.03$, 
$\Delta$log $r_{h} / \Delta M_V = 0.00 \pm 0.03$, 
$\Delta c / \Delta M_V = -0.07 \pm 0.04$, and
$\Delta \epsilon/ \Delta M_V = 0.02 \pm 0.01$. 
The scale radii $r_c, r_{h}$ and ellipticity 
display considerable scatter and no significant trends
with cluster luminosity, while the central concentration
increases weakly with luminosity;  
weak correlations of $r_c$ and $c$ vs. luminosity were
derived by \citet{mcl00}  for the Milky Way clusters (shown in Figure \ref{mv_corr}
by the dashed lines).  The lack of correlation of the characteristic
radii $r_c, r_h$ with luminosity is closely connected to the narrow distribution
of clusters in the fundamental plane, which we discuss in the next section.
Some of the $r_c$-distribution differences between the two galaxies 
are likely to be due simply to the sample bias we noted above.
For the $c(L)$ graph (lower left panel) we note 
that the higher-resolution, higher-S/N data (solid dots)
follow the Milky Way relation much more closely than do the 
lower-resolution data (and would do so even more closely if the
one ``high'' point at upper left were ignored, due to the very faint C102),
suggesting perhaps that the $c-$values for the
inner-halo data are systematically underestimated.  The higher half-light
radii from the lower-resolution data
(upper right panel) may also be an artifact of the same effect.
Higher-resolution data for more clusters will be needed before these
concerns can be cleared up definitively.

In Figure \ref{rgc_corr} the same four quantities are plotted
against projected galactocentric distance $R_{gc}$.
The two subgroups of measured clusters (WFPC2 vs.~STIS) are more clearly 
separated in Figure \ref{rgc_corr},
because almost all of the lower-resolution WF data come from the inner-halo
clusters.  The same net offset towards lower $c$ at smaller $R_{gc}$
which we noted above 
is clearly visible in the lower left panel.  Physically,
we would have expected a higher mean $c$ closer to the
center of the galaxy because of the stronger effects of dynamical evolution,
as well as a smaller mean $r_h$ (see below), again indicating that the low
WF resolution may have compromised the model fits for the inner clusters.
On the other hand, any subtle real trends with $R_{gc}$ may be washed out
by the fact that we are observing only the projected distance rather than
the true three-dimensional galactocentric distance.

As a further test of the effects of image resolution on the
deduced structural parameters \citep[see][for a more extensive
discussion including simulations]{hol99}, we carried out some
brief numerical trials with the STIS data.  
For four clusters (C22, C23, C102, C103), we rebinned the original images
down to $2 \times$ lower resolution ($0\farcs1$ per binned pixel)
to roughly simulate the resolution of the WF camera.  We then
remeasured the cluster profiles and fit the 1-D King profiles
as before.  The results for the key structural parameters
$(W_0, c, r_c, r_h)$ were then compared with the original profiles. 
C23 is among the brightest in our sample,
while C22 and C103 are of intermediate brightness and C102 is among
the faintest.  For the three brightest,
the fitted $W_0$ values agreed with their higher-resolution 
counterparts to within 5\%.  The $r_c$ and $c$ values tended to
be overestimated on the low-resolution images by about
5\%-10\%, while $r_h$ was overestimated by about 20\%.
For the very faint C102, the agreement with the high-resolution
parameters was no better than $\pm$50\%.  These
trials suggest to us that the $2 \times$ lower resolution by itself does not seriously
damage the ability to measure the structural parameters (which at best are
internally uncertain at the 10\% level; see above).  Instead, 
the signal-to-noise level of the data (exposure time, cluster luminosity,
background light)
seems to be a more important factor at the levels we are dealing with,  
and as noted above, the \citet{hol99} WFPC2 data were taken with 
shorter exposure times and
higher background light from the inner halo.

\subsection{Metallicity and Characteristic Cluster Size}

Finally, in Figure \ref{feh_rh} we show the scatter plot of
cluster half-light radius versus measured metallicity.  Since $r_h$
is closely related to the half-mass radius $r_{hm}$, which remains
relatively constant over long timescales of dynamical evolution,
$r_h$ represents a useful characteristic scale size for the cluster. A formal
least-squares solution gives $r_h$(arcsec) 
$= (0.381 \pm 0.010) + (0.064 \pm 0.067)$ [Fe/H],
confirming the immediate visual impression that no significant correlation
exists.  Dividing the sample more or less arbitrarily into metal-poor
([Fe/H] $< -1$) and metal-rich ([Fe/H] $> -1$) groups, we find that the mean
characteristic radii $\langle r_h \rangle$ are $(7.37 \pm 1.03)$ pc
and $(7.14 \pm 0.76)$ pc, indistinguishably different.

This result is formally at odds with what has been found for the M31 clusters
\citep{bar02} and for the clusters in certain giant E galaxies
\citep{kun99,lars01} and in the giant Sa galaxy M104 \citep{lar01a}, 
where mild systematic {\sl decreases} in cluster
size with increasing metallicity have been claimed to exist.  
Our result, however, may be a simple consequence of working with
a small sample which is selection-biased towards large clusters.
A much more comprehensive imaging set will be needed to explore this
potentially important correlation adequately.

\section{Binding Energy and the Fundamental Plane}

\citet{djo95} demonstrated that in the trivariate space of
central surface brightness, velocity dispersion, and core radius,
the Milky Way globular clusters occupy only a relatively narrow region now called
the ``fundamental plane'', similar to that expected if the cores
were virialized structures.  \citet{bar02} have shown definitively
with a comparably large sample of M31 clusters that they too fall in
very much the same restricted region of parameter space.  
Lacking any extensive direct measures of the
internal velocity dispersions for the NGC 5128 clusters, we cannot
define the same graphs as for the Milky Way and M31. 
We have, however, carried out a series of consistency tests which strongly
suggest to us that the clusters in this giant E galaxy define a
closely similar FP.

Here we cast the discussion in the way laid out by \cite{mcl00},
in terms of the cluster {\sl binding energies} $E_b$.
The binding energy is fundamentally related
to the cluster {\sl mass} $M$, and
on basic virial-theorem grounds, we expect a dependence of the form
$E_b \sim f(c) \cdot M^2/r_c$.  
Combining McLaughlin's equations (A6) and (A7),
we obtain $E_b$ in terms of the cluster core radius, luminosity, 
central concentration, and mass-to-light ratio:
\begin{equation}
E_b \, = \, G \Big({4 \pi \over 9} \Upsilon \Big)^2 \, {L^2 \over r_c} \, {{\cal E}(c) \over {\cal L}(c)^2}
\end{equation}
where $\cal E, \cal L$ are dimensionless functions 
of $c$, and $\Upsilon$ is the mass-to-light ratio.  

Independent calibrations of $\Upsilon$ require direct spectroscopic 
measurements of their internal velocity dispersions $\sigma_v$.  
As yet, little such material is available for NGC 5128
by comparison with the Milky Way and M31.
Velocity dispersion measurements for 10 of the brightest NGC 5128
clusters are described by \citet{dub94} and stated to be in the range $15 - 25$ km s$^{-1}$.
A subsequent discussion by \citet{dub97} uses these to show that the
NGC 5128 clusters fall, within the measurement uncertainties, on the top end
of the same FP as in
the Milky Way and M31.  The list of measured velocities from Dubath
(private communication, and in preparation) 
shows that there are four clusters in common 
between his sample and ours, and for these we can carry out complete calculations
of $\Upsilon_V$.  These four objects are listed in Table 4.
We combine McLaughlin's (2000a) expression (A1) defining $\Upsilon$ with his
equation (A6) to eliminate the central luminosity density $j_0$, giving
the mass-to-light ratio in terms of our measured parameters,
\begin{equation}
\Upsilon \, = \, {9 {\cal L}(c) \, \sigma_0^2 \, r_c \over {4 \pi G L}}
\end{equation}
where the scale velocity $\sigma_0$ is given as a ratio of the measured
dispersion $\sigma_v$ by McLaughlin's relation (B1);
for any $c-$values in our range of interest, we have $\sigma_0 \simeq \sigma_v$
to within a few percent.  The structure of the cluster (particularly,
its central concentration $c$) enters through the dimensionless function
${\cal L}$.  Since ${\cal L}$ varies by more than a factor of 20
over the $c-$range occupied by typical globular clusters, 
$\Upsilon$ is quite sensitive to both the velocity dispersion
and the cluster structure; we need to know both to estimate the
cluster masses correctly.

Table 4 lists the measured velocity dispersions from Dubath;
the structural luminosity parameter ${\cal L}$; and the calculated
mass-to-light ratio $\Upsilon_V$.  Taking into account the measurement
uncertainties in $\sigma_v$, $c$, $r_c$, the integrated $V$ magnitudes,
and the galaxy distance itself, we find that $\Upsilon$ is uncertain to
$\pm30$\% for a given object.  The weighted average over the four clusters is
$\langle \Upsilon_V \rangle = 1.56 \pm 0.24$.  By comparison, 
McLaughlin's mean value $\langle \Upsilon_V\rangle = 1.45 \pm 0.1$ 
(a mean over 39 Milky Way clusters) is not significantly different,
and provides an encouraging consistency test that the two groups of
clusters are similar.

If we now 
{\sl assume} $\Upsilon_V = 1.45$ (the more precise Milky Way mean)
then numerically Eq.~(2) becomes
\begin{equation}
{\rm log } E_b \,{\rm (ergs)} \, = \,  41.545 \, + \, 2 {\rm log } (L/L_{\odot})
\, - \, {\rm log } (r_c/{\rm pc}) \, + \, {\rm log } {\cal E}(c) \, - 
\, 2 {\rm log } {\cal L}(c) \, .
\end{equation}
Given $(L, r_c, c)$ for each cluster we can then estimate its binding energy
{\sl modulo} our uncertainty in the mass-to-light ratio.

The results for our combined sample of 43 clusters are displayed graphically
in Figure \ref{ebplane} as $E_b$ versus $L$.  
A straightforward linear least-squares correlation, unweighted, gives
\begin{equation}
{\rm log } \, E_b \,{\rm (ergs)} \, = \, (40.41 \pm 0.40) \, + \, \
(1.952 \pm 0.070) \, {\rm log} \, (L/L_{\odot}) \, \, {\rm (NGC~5128)}
\end{equation}
with a residual scatter in log $E_b$ of $\pm 0.22$ dex. 
The measurement uncertainties in the 
quantities used to calculate $E_b$ generate
expected scatters of $\pm 0.04$ in 
log $r_c$, $\pm 0.08$ in log ${\cal E}$, and $\pm 0.09$ in
log ${\cal L}$.  Adding these in quadrature, we obtain a net expected
dispersion of $\pm 0.20$ in log $E_b$, sufficient to explain
almost all of the observed spread in Figure \ref{ebplane}.

For comparison, from a sample of 109 non-core-collapsed Milky Way clusters
\citet{mcl00} obtained the correlation (see his Figure 6 and accompanying text)
\begin{equation}
{\rm log } \, E_b \,{\rm (ergs)} \, = \, (39.89 \pm 0.38) \, + \, \
(2.05 \pm 0.08) \, {\rm log} \, (L/L_{\odot}) \, \, {\rm (Milky~Way)}.
\end{equation}
The dispersion about this latter relation is $\pm 0.53$ dex, though it should
be noted that the scatter is $\sim$30\% smaller if $E_b$ is
normalized to a constant Galactocentric distance; see the
discussion of McLaughlin and his Figure 12.

A basic $L^2$ proportionality of $E_b$ is of course built directly into
Eq.~(2).  What is more interesting is that the observed slope of the actual
sequence is so close to 2.00, and the scatter around the relation
is so small, that the other factors
($r_c, \Upsilon, {\cal L}, {\cal E}$) {\sl in combination} must not vary importantly with cluster mass
$M$ itself \citep[see][for more extensive discussion]{mcl00,bar02}.
Secondly, the NGC 5128 sequence falls along a locus which, within its
own uncertainties, is indistinguishable from the Milky Way sequence, 
consistent with the claim 
that we are looking at very much the same type of object with the same
mean mass-to-light ratio (see below).  

The NGC 5128 results also demonstrate clearly that the $E_b(L)$
relation continues upward along the same
slope to significantly higher cluster luminosities (masses)
than were previously observed.  Whereas the Milky Way has only two
clusters more luminous than $M_V = -10$, our NGC 5128 sample has 14, and
the relation now extends to almost a factor of three higher,
reaching an equivalent mass of $4 \times 10^6 M_{\odot}$ for $\Upsilon = 1.45$.

Small though it is, some of the residual scatter in Figure \ref{ebplane} may
correlate with other external factors such as age, metallicity, or
galactocentric distance.  Although we cannot evaluate age directly at present,
the residual correlations against distance and metallicity are shown
in Figure \ref{eb_resid}.  
To check for metallicity effects, we use 16 clusters in our list for
which [Fe/H] values are known from
the Washington photometry by \citet{har92}.  Although $(V-I)$ colors are
available for the other clusters in the list,
most of these are inner-halo objects where some of the $(V-I)$ values are
confused with possible internal reddening differences and
are, in any event, not very sensitive to metallicity \citep[see][]{hol99}.
The resulting correlation (lower panel of Figure \ref{eb_resid}) is
$\Delta$ log $ (E_b/L^2) / \Delta$ [Fe/H] $= (0.00 \pm 0.07)$.
Just as for the Milky Way, we find no trace of any dependence of
binding energy on metallicity.

On the other hand, a change in mean $E_b$ with galactocentric distance is expected.
We have $E_b \sim M^2/r_{hm}$ by definition, and 
at least for the Milky Way, the characteristic cluster size $r_{hm}$ 
is known to vary with $R_{gc}$.  \citet{mcl00} finds 
empirically that $r_{hm} \sim R_{gc}^{0.4}$ for the Milky Way system,
consistent with rough theoretical arguments that predict 
$r_{hm} \sim R_{gc}^{0.5}$ \citep[e.g.][]{har94} for an isothermal
halo potential well. That is, the linear scale sizes of the clusters
should reflect the sizes of their initial protocluster gas clouds, and
the clouds will have larger scale sizes in the outer regions 
of the halo where they are under lower ambient pressures.
In the NGC 5128 data sample we can look for the same effect, although
(unlike the Milky Way) we can work only with the {\sl projected}
galactocentric distances rather than the full three-dimensional $R_{gc}$.
In the upper panel of Figure \ref{eb_resid} we show $(E_b/L^2)$ versus
$R_{gc}$, where the latter now denotes the projected distance.
The resulting correlation is
$\Delta$ log $ (E_b/L^2) / \Delta$ log $ R_{gc} = (-0.325 \pm 0.069)$.
This is a significant trend in the same sense as was found for the Milky Way, though
projection into two dimensions has, as expected, left
a shallower slope.\footnote{We note in passing that the trend 
shown in Figure \ref{eb_resid}, when it is
extended to a larger and more well determined statistical sample,
may eventually place some interesting new constraints on the formation history
of NGC 5128.  The reason is that the clusters which are {\sl now} in the
halo have characteristic radii $r_{hm}$ that still reflect their
place of formation.  For example, if they formed within disks of large
progenitor galaxies under conditions of high ambient pressure, they would
maintain their original small characteristic sizes even after the disks
merged to produce NGC 5128 and projected much of their material well out into
the halo.  On the other hand, if the clusters formed within lower-mass
dwarfs which later merged, they would have larger effective radii
consistent with the weak potential wells in which they formed.
Detailed simulations will be needed to test whether 
the slope of a relation such as in Figure \ref{eb_resid} can be maintained
after a long series of mergers of many types of progenitor galaxies.}

Too much reliance cannot be placed on this latter result, since the datapoints
at small $R_{gc}$ may have measurement
biasses as discussed above.  Nevertheless, if we now rather boldly accept the
main trend of the effect and normalize all the $E_b$ values to a
fiducial distance (10 kpc) by defining
\begin{equation}
E_b^{\star} \, = \, E_b \cdot (R_{gc}/10)^{0.325} \, \, ,
\end{equation}
the net result is to yield a corrected binding-energy relation as 
shown in Figure \ref{ebstar} with the equation
\begin{equation}
E_b^{\star} \, = \, (40.07 \pm 0.32) \, + \, (2.00 \pm 0.06) \, {\rm log} (L/L_{\odot})
\end{equation}
and with a dispersion of only $\pm 0.18$ dex over its entire run.

The analysis so far has assumed a uniform age for all clusters in the sample.
Age differences would be expected 
to enter mainly through the mass-to-light ratio,
which decreases with increasing age.
Since $E_b(L) \sim \Upsilon^2$, any mean age difference between the Milky Way
and NGC 5128 would then show up
in Figure \ref{ebplane} as a net offset of the data points away
from the Milky Way line.  If, for example, the NGC 5128 clusters were actually much
younger, the points in Figure \ref{ebstar}
would then have fallen above the Milky Way line:  that is, by adopting the
mean $\Upsilon = 1.45$ valid for old clusters, we would have overestimated
their masses (and thus their binding energies) for a given luminosity.
Approximate predictions for the amount of the effect are shown
in Figure \ref{ebstar} by the two dotted lines parallel to the main
relation, the upper one for a cluster age of 2 Gy and the lower one
for 5 Gy.  We have used Bruzual/Charlot relations \citep[quoted by][]{whi97}
for the change in cluster luminosity $\Delta M_V$ with age to estimate these
offsets.  Thus for {\sl individual} clusters, age differences of 
factors of two are discernible on this diagnostic graph.
In practice, no offset from the Milky Way cluster line is detectable to well
within the observational scatter, so our data
are strongly consistent with the claim that we are looking at basically the
same type of object in both galaxies:  old, luminous star clusters with 
similar mass-to-light ratios.

\citet{mcl00,mcl00b} has suggested that the three quantities
$(L, E_b^{\star}, c)$ provide a physically transparent way to describe the
fundamental plane or ``$\epsilon-$space'' for globular cluster structures.
In Figure \ref{ebstar} we are essentially
looking at the FP edge-on, but tilted.  A rectified form of the FP can be generated 
from  a suitable rotation of the $(L, E_b^{\star})$ plane, given by
$\epsilon_1 \equiv$ log $E_b^{\star} - 2$ log $L$,
$\epsilon_2 \equiv 2.00 {\rm log} E_b^{\star} + {\rm log} L$, and
$\epsilon_3 \equiv c$.  The edge-on view (Figure \ref{ebstar}) is essentially
a rotated version of ($\epsilon_1, \epsilon_2)$.
A fully ``face-on'' view of the FP can then be constructed
from a plot of $\epsilon_2$ against $\epsilon_3$.
This plot is shown in Figure \ref{fp}.  
There is little left in this plane but
pure scatter, but this may simply be reflecting
the limitations of our small sample.
A larger sample extending downward to much lower luminosities
might reveal the same overall trend with $c$ (lower central concentration
at lower luminosity) that is clearly found in the Milky Way
\citep[compare Figure 13 of][shown here as the dashed line]{mcl00}.

The remarkably tight $E_b(L)$ relation, for clusters at all galactocentric
distances, hints that it was set largely by the cluster formation
process \citep{mcl00,mcl00b}.  For protocluster gas
clouds $E_b$ should behave in proportion to $M^{1.5} R_{gc}^{-0.5}$, if they
are constrained at the time of star formation by an external pressure
$P_s$ which itself varies as $P_s \sim R_{gc}^{-2}$ in the isothermal
potential well of the protogalaxy \citep[see][]{mcl00b,har94}.  
\citet{mcl00b} gives the expected dependence of $E_b$(gas cloud) for
the case of the Milky Way halo with circular velocity $V_c = 220$ km s$^{-1}$.
Renormalizing to NGC 5128 with $V_c = 245$ km s$^{-1}$ \citep{hui95}, 
we find under the same assumptions
$E_b$(gas) $\simeq 4.76 \times 10^{42}$ erg $\cdot M^{1.5} \cdot
(R_{gc}/10 {\rm kpc})^{-0.5}$.  The location of this line, for 
$R = 10$ kpc and $(M/L)=1.45$ as before, is shown in Figure \ref{ebstar}.
The fully formed clusters we observe today obey a 
distinctly different and steeper relation ($E_b \sim M^2$ rather than
$M^{1.5}$), indicating perhaps that the lower-mass clusters lost relatively more
of their binding energy through early gas loss. 

Under the assumptions of this doubtless-oversimplified model, the 
protocluster $E_b$ line should represent an upper boundary
which could only be reached by the actual star clusters if their star
formation efficiency approached 100\%.  However, we see that at the upper end
the most massive known clusters lie {\sl above} the gas-cloud relation
by as much as a factor of two.
At this point we can only speculate on possible interpretations:
our schematic model for the gas clouds may be too rough, or the clusters
may have evolved dynamically away from their initial conditions to a state
of higher binding energy.  The dynamics of cloud collapse and star
formation during a cluster's earliest stages are poorly understood,
and there may be various mass-dependent phenomena at work whose results
we see in the present-day $E_b(L)$ correlation.

\section{Summary}

We have used new imaging data from the HST STIS and WFPC2 cameras
to derive structural parameters for globular
clusters in the halo of the giant elliptical galaxy NGC 5128.
We find that classic, single-mass King models describe their
observed light profiles extremely well, allowing us to derive
parameters $(r_c, r_h, c, \epsilon, L)$ for direct comparison with
the globular clusters in other galaxies.  The NGC 5128 clusters occupy very much
the same regions of parameter space as those in the Milky Way, with the exception
that they have a higher range of ellipticities:  they occupy the
range $0 < \epsilon < 0.3$ more or less uniformly, and among various comparison
galaxies within the Local Group, we find that they most nearly
resemble the old clusters in the LMC and M31 in this respect.
We also find half a dozen
luminous clusters which may have ``extratidal light'' which is
possibly due to active tidal stripping or residual
field-star populations from disrupted dwarf satellite galaxies, but
may also be the signature of anisotropic velocity distributions.

Lastly, we find that the NGC 5128 clusters delineate a relation
between binding energy $E_b$ and luminosity $L$ which is even
tighter than in the Milky Way and in exactly the same region
of the ``fundamental plane''.  This work provides additional
evidence that globular cluster formation processes were remarkably
similar in galaxies of very different types.

Considerable further progress can be made in understanding
the structures of clusters in this keystone galaxy if we can
obtain a more extensive sample of objects over a wide range
of galactocentric distances and at the highest resolution possible.
In addition, direct spectroscopic measurements of their
velocity dispersions are needed to check the key assumptions we
have made about their mass-to-light ratios.

\acknowledgments

This work was supported by the Natural Sciences and Engineering
Research Council of Canada through research grants to the authors.
WEH and GLHH are pleased to acknowledge the hospitality and support at 
Mount Stromlo Observatory (RSAA/ANU)
during research leaves when this paper was written.  STH acknowledges
support from NASA grant NAG5-9364.  An anonymous referee provided
several constructive suggestions which improved the manuscript.
We also thank Georges Meylan for a helpful comment about the 
applicability of anisotropic velocity distributions.
Finally, we are extremely grateful
to Pierre Dubath for transmitting his cluster velocity
dispersion data in advance of publication, allowing us to make crucial
consistency checks on the mass-to-light ratios.


\clearpage


\begin{deluxetable}{lcccccrc}
\tablecaption{List of Individually Imaged NGC 5128 Globular Clusters \label{coords}}
\tablewidth{0pt}
\tablehead{
\colhead{Cluster ID} & \colhead{$\alpha$ (J2000)} & \colhead{$\delta$ (J2000)}
& \colhead{$R_{gc}$ (kpc)} & \colhead{$V$} & \colhead{$(V-I)$} & \colhead{$M_V$}
& \colhead{detector}\\
}
\startdata
 C40  & 13 23 42.29 & -43 09 39.7 &  24.50 &   18.878 &         &  -9.47 &  stis \\
 C41  & 13 24 38.92 & -43 20 08.0 &  24.40 &   18.556 &         &  -9.79 &  stis \\
 C29  & 13 24 40.35 & -43 18 06.3 &  22.15 &   17.936 &         & -10.41 &  stis \\
 G19  & 13 24 46.35 & -43 04 12.6 &   9.47 &   19.069 &   1.189 &  -9.28 &  wf3  \\
 G277 & 13 24 47.29 & -42 58 32.2 &   9.10 &   19.030 &         &  -9.32 &  stis \\
 C2   & 13 24 51.49 & -43 12 12.2 &  14.99 &   18.334 &         & -10.02 &  stis \\
 C100 & 13 24 51.80 & -43 04 33.7 &   8.57 &   19.85: &         &  -8.50 &  stis \\
 C100 & 13 24 51.92 & -43 04 33.8 &   8.57 &   20.08: &   1.277 &  -8.27 &  pc1  \\
 G302 & 13 24 53.07 & -43 04 35.9 &   8.36 &   19.156 &         &  -9.19 &  stis \\
 G302 & 13 24 53.19 & -43 04 35.9 &   8.36 &   19.103 &   1.295 &  -9.25 &  pc1  \\
 C11  & 13 24 54.80 & -43 01 22.7 &   6.99 &   17.695 &         & -10.66 &  stis \\
 C31  & 13 24 57.52 & -43 01 09.1 &   6.40 &   18.366 &         &  -9.98 &  stis \\
 C32  & 13 25 03.30 & -42 50 47.2 &  13.12 &   18.308 &         & -10.04 &  stis \\
 C44  & 13 25 31.60 & -43 19 24.3 &  21.26 &   18.605 &   1.163 &  -9.75 &  pc1  \\
 C17  & 13 25 39.63 & -42 56 00.7 &   6.50 &   17.612 &         & -10.74 &  stis \\
 C101 & 13 25 40.47 & -42 56 02.7 &   6.53 &   20.342 &         &  -8.01 &  stis \\
 C102 & 13 25 52.07 & -42 59 14.4 &   5.65 &   21.431 &         &  -6.92 &  stis \\
 C21  & 13 25 52.70 & -43 05 48.1 &   7.60 &   17.769 &         & -10.58 &  stis \\
 C22  & 13 25 53.54 & -42 59 09.0 &   5.98 &   18.143 &         & -10.21 &  stis \\
 C23  & 13 25 54.55 & -42 59 26.8 &   6.06 &   17.191 &         & -11.16 &  stis \\
 C103 & 13 25 54.98 & -42 59 15.4 &   6.22 &   18.880 &         &  -9.47 &  stis \\
 C104 & 13 25 59.43 & -42 55 32.2 &   9.40 &   19.957 &         &  -8.39 &  stis \\
 G221 & 13 26 01.06 & -42 55 14.8 &   9.88 &   19.203 &         &  -9.15 &  stis \\
 C25  & 13 26 02.79 & -42 56 58.3 &   8.92 &   18.333 &         & -10.02 &  stis \\
 G293 & 13 26 04.27 & -42 55 45.4 &  10.00 &   19.122 &         &  -9.23 &  stis \\
 C105 & 13 26 05.12 & -42 55 37.0 &  10.25 &   22.006 &         &  -6.34 &  stis \\
 C7   & 13 26 05.35 & -42 56 33.7 &   9.64 &   17.104 &         & -11.25 &  stis \\
 C106 & 13 26 06.15 & -42 56 45.4 &   9.66 &   21.280 &         &  -7.07 &  stis \\
 C37  & 13 26 10.53 & -42 53 44.0 &  12.56 &   18.342 &         & -10.01 &  stis \\
\enddata

\end{deluxetable}

\clearpage

\begin{deluxetable}{lcccccccc}
\tablecaption{Measured Structural Parameters \label{parameters}}
\tablewidth{0pt}
\tablehead{
\colhead{Cluster} & \colhead{$W_0$} &
\colhead{$r_c$ (arcsec)} & \colhead{$r_{h}$ (arcsec)}
& \colhead{$c$} & \colhead{$\mu_V^0$} & \colhead{$\epsilon$} 
& \colhead{$d\epsilon \over dr$} & \colhead{$\theta$} \\
}
\startdata
C40  & 7.5 &  0.108 &  0.570 &  1.69 & 17.54 & 0.17 & + & 74$^o$ \\
C41  & 8.1 &  0.042 &  0.322 &  1.87 & 15.91 & 0.05 & + & 174: \\
C29  & 8.1 &  0.064 &  0.492 &  1.87 & 16.00 & 0.11 & + & 92 \\
G19  & 7.4 &  0.101 &  0.489 &  1.65 & 17.37 & 0.20 && 173  \\
G277 & 6.9 &  0.067 &  0.248 &  1.49 & 16.53 & 0.05 && 160: \\
C2   & 8.5 &  0.043 &  0.456 &  1.99 & 16.10 & 0.04 && 82: \\
C100(st) & 8.0 &  0.079 &  0.563 &  1.86 & 18.22 & 0.10 && 188 \\
C100(pc) & 7.5 &  0.093 &  0.464 &  1.67 & 18.04 & 0.11 && 160 \\
G302(st) & 7.2 &  0.052 &  0.231 &  1.60 & 16.38 & 0.14 && 163 \\
G302(pc) & 7.1 &  0.056 &  0.230 &  1.56 & 15.98 & 0.09 && 163 \\
C11  & 8.2 &  0.070 &  0.559 &  1.88 & 15.97 & 0.26 & $-$ & 165 \\
C31  & 7.7 &  0.049 &  0.284 &  1.74 & 15.70 & 0.10 & $-$ & 120 \\
C32  & 8.8 &  0.030 &  0.389 &  2.06 & 15.75 & 0.06 && 24: \\
C44  & 7.0 &  0.067 &  0.408 &  1.70 & 16.36 & 0.06 && 160 \\
C17  & 6.6 &  0.122 &  0.408 &  1.43 & 16.12 & 0.07 && 116 \\
C101 & 6.5 &  0.110 &  0.345 &  1.38 & 18.60 & 0.09 && 156 \\
C102 & 9.2 &  0.060 &  1.112 &  2.18 & 20.36 & 0.21 && 72 \\
C21  & 8.1 &  0.065 &  0.498 &  1.86 & 15.87 & 0.33 && 33 \\
C22  & 7.3 &  0.059 &  0.272 &  1.62 & 15.59 & 0.09 & + & 82 \\
C23  & 7.5 &  0.047 &  0.237 &  1.67 & 14.34 & 0.14 & &60 \\
C103 & 7.6 &  0.053 &  0.288 &  1.71 & 16.29 & 0.15 & + & ~~2\\
C104 & 7.2 &  0.056 &  0.239 &  1.58 & 17.27 & 0.16 && 57 \\
G221 & 7.1 &  0.060 &  0.246 &  1.56 & 16.59 & 0.07 && 137 \\
C25  & 8.2 &  0.053 &  0.447 &  1.90 & 16.18 & 0.13 & $-$ & 100 \\
G293 & 7.8 &  0.036 &  0.222 &  1.76 & 16.03 & 0.05 & & 88 \\
C105 & 7.2 &  0.191 &  0.809 &  1.57 & 21.60 & 0.18 && 20 \\
C7   & 8.0 &  0.076 &  0.537 &  1.83 & 15.39 & 0.13 && 15 \\
C106 & 6.8 &  0.035 &  0.124 &  1.46 & 17.70 & 0.11 && 25 \\
C37  & 8.1 &  0.031 &  0.238 &  1.87 & 15.22 & 0.02 && 135: \\
\\
$\pm$ & 0.6 & 0.006 &  0.010 &  0.15 & ~0.20 & 0.07 && 10$^o$ \\
 \enddata

\end{deluxetable}

\clearpage 

\begin{deluxetable}{ccccc}
\tablecaption{Clusters With Possible Extratidal Light \label{xtl.tab}}
\tablewidth{0pt}
\tablehead{
\colhead{Cluster ID} & \colhead{$R_{gc}$ (kpc)} & \colhead{$M_V$} & 
\colhead{$\Delta b$ (adu)} & \colhead{Fraction in XT Light} \\
}
\startdata
 C7   & 9.64 & $-11.25$ & 12 &$0.11 \pm 0.05$ \\
 C23  & 6.06 & $-11.16$ & 15 & $0.07 \pm 0.04$ \\
 C25  & 8.92 & $-10.02$ &  9 & $0.15 \pm 0.15$ \\
 C29  & 22.15 & $-10.42$ & 12 & $0.17 \pm 0.08$ \\
 C32  & 13.12 & $-10.04$ & 10 & $0.15 \pm 0.11$ \\
 C37  & 12.56 & $-10.01$ & 10 & $0.07 \pm 0.09$ \\
\enddata

\end{deluxetable}
 
\begin{deluxetable}{cccc}
\tablecaption{Estimates of Mass-to-Light Ratios \label{ml.tab}}
\tablewidth{0pt}
\tablehead{
\colhead{Cluster ID} & \colhead{$\sigma_v$ (km s$^{-1}$)} & 
\colhead{${\cal L}(c)$} & \colhead{$\Upsilon_V$ ($M_{\odot}/L_{\odot}$)} \\
}
\startdata
 C7   & $19.8 \pm 1.3$ & 39.56 & $1.45 \pm 0.44$ \\
 C17  & $20.9 \pm 1.6$ & 21.71 & $2.38 \pm 0.71$ \\
 C21  & $16.3 \pm 2.1$ & 41.69 & $1.65 \pm 0.50$ \\
 C23  & $26.1 \pm 1.5$ & 30.59 & $1.32 \pm 0.40$ \\
\enddata

\end{deluxetable}

\clearpage

\begin{figure}
\caption{Sample sections of two STIS fields, both $30''$ (600 px) on a side.
{\sl Left panel:} The very luminous NGC 5128 
outer-halo cluster C29 is at lower right, with a bright star at 
upper left.  C29 has an extended envelope and a
noticeably elliptical structure, $\epsilon \simeq 0.1$.
{\sl Right panel:} The bright cluster C17 is at top, a fainter
cluster C101 at center, and a bright star at bottom, with other faint stars
toward the right side.  Notice the increased ``graininess'' of the background
in the right panel compared with the left, which is due to the semi-resolved
population of red giant stars in the halo of NGC 5128.  C17 and C101 are 
at projected distances of only 6.5 kpc from the galaxy center, whereas C29
is 22 kpc from the center and thus has much less surrounding light from the
faint halo stars. Note also that the two inner-halo clusters have more compact
structures.
\label{samples}}
\end{figure}

\clearpage
\begin{figure}
\plotone{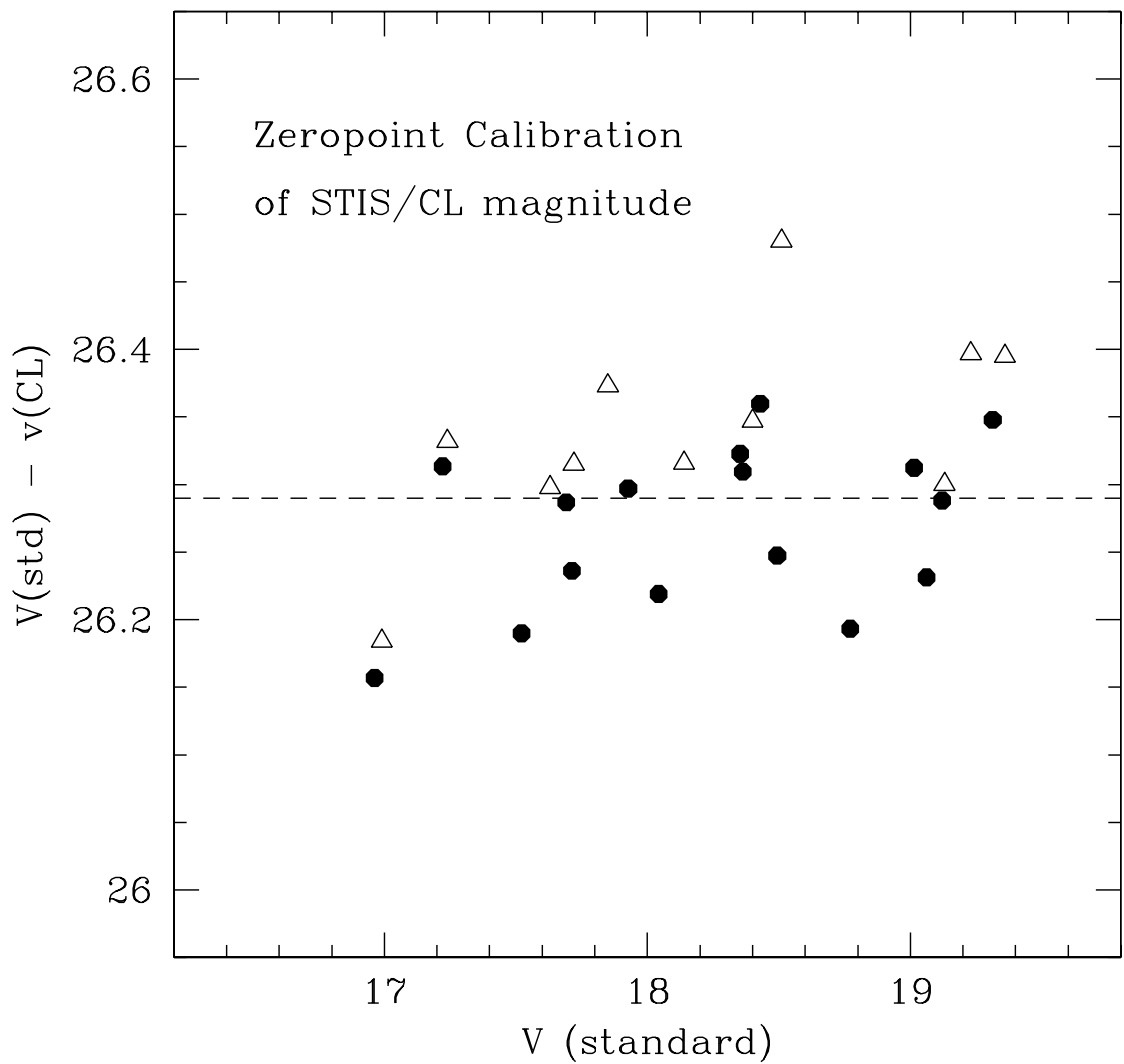}
\caption{Photometric calibration of the STIS/50CCD ``CL'' magnitudes.
Here the difference between the standard $V$ magnitude and
instrumental $v(CL)$ is plotted against $V$.  The open triangles represent
objects from the Tonry \& Schechter (1990) $(V,V-I)$ data, and the
filled circles are data from G.Harris et al.~(1992)
converted from the Washington system indices into $V$ through
Eq.~(1).  The dashed line is at our adopted mean of 26.29.
\label{Vcalib}}
\end{figure}

\clearpage
\begin{figure}
\plotone{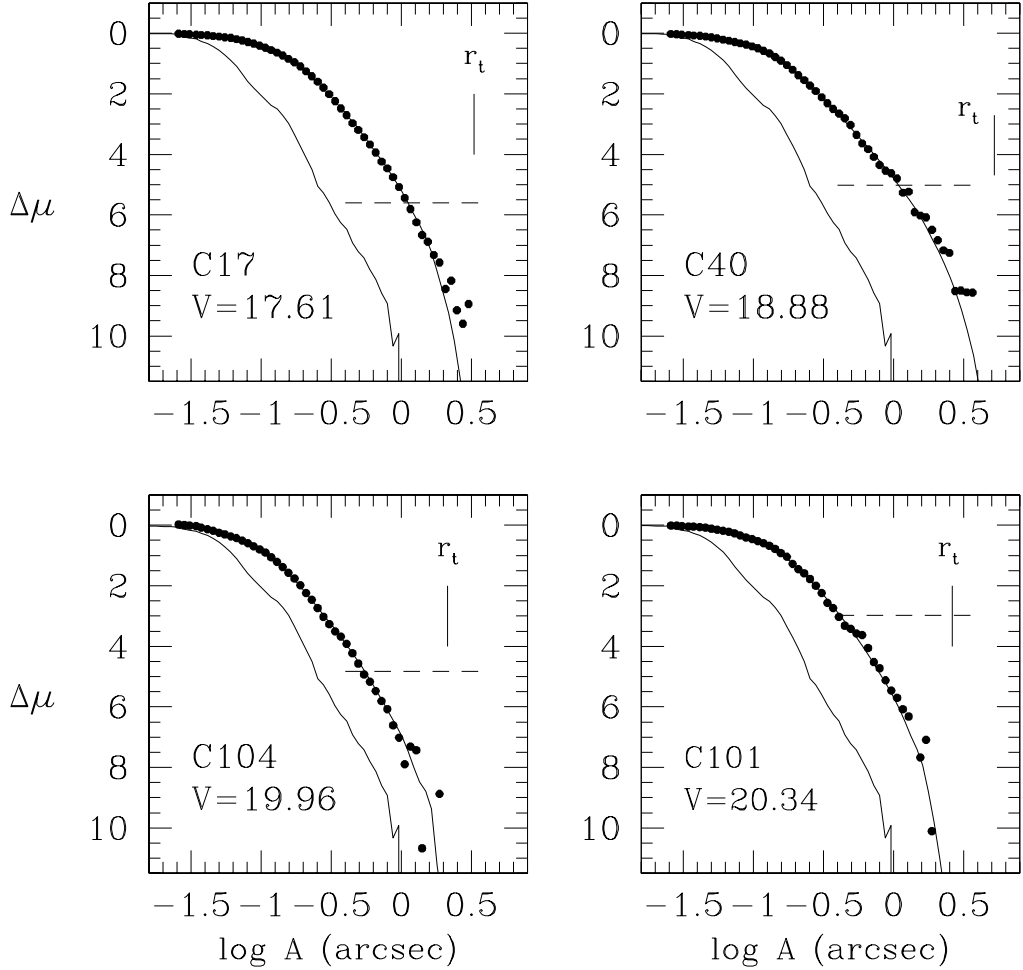}
\caption{Typical profiles for four clusters imaged with
the STIS camera in unfiltered 50CCD mode.  Here $\Delta \mu$
is the surface intensity in $V$ in magnitudes per unit area
relative to the central surface intensity of the cluster,
and $A$ in arcseconds is the semimajor axis. In each panel,
the leftmost solid line is a PSF profile for starlike
objects, and the horizontal dashed line indicates the local
background light intensity.  Solid dots indicate the surface
brightness profile determined from the {\sl stsdas.ellipse} code,
while the solid line through each data set is the best-fitting
King model convolved with the PSF profile.
The nominal tidal radius $r_t$ is indicated by the short vertical
line at the right edge of each panel.
\label{profiles}}
\end{figure}

\clearpage

\begin{figure}
\plotone{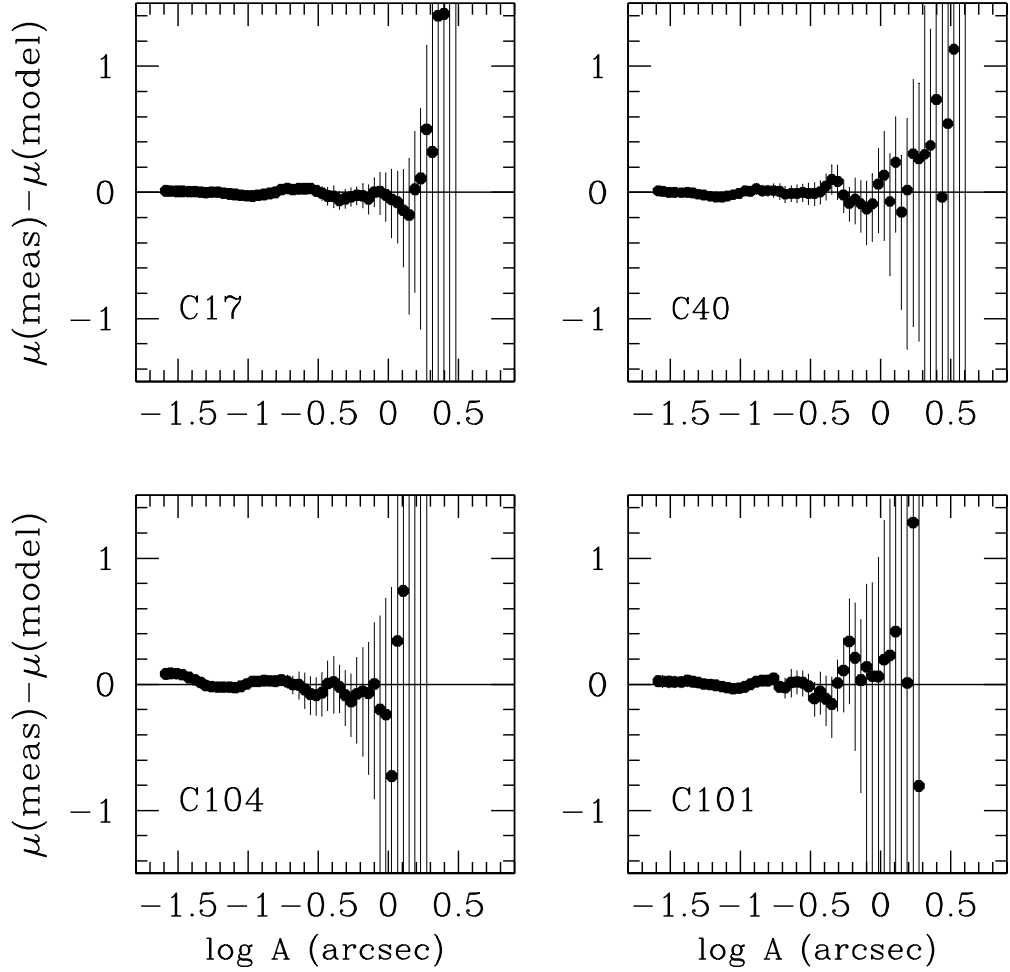}
\caption{Residual light profiles (data minus model curve) for
the four clusters in the previous figure.  The dots represent the
magnitude difference between the measured surface intensity and
the best-fit King model, while the error bars show the internal
uncertainty of each data point.  For $\mu-$values that fall a magnitude
or more fainter than the background intensity, the uncertainties
rapidly blow up and exert no constraints on the fit.
\label{resid}}
\end{figure}

\clearpage
\begin{figure}
\plotone{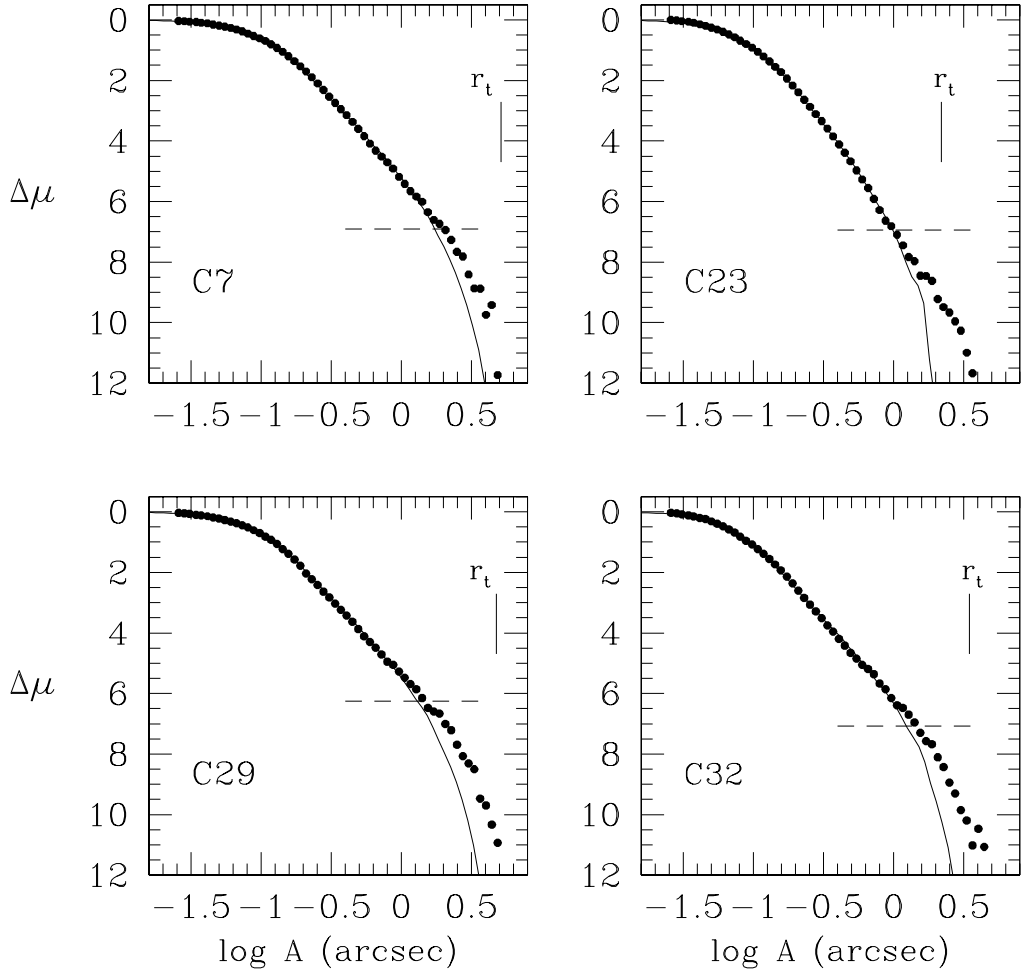}
\caption{Light profiles for four clusters with possible
``extratidal light'' at large radii.  In each panel the
thin solid line represents the best-fitting King model,
while the points show the actual cluster profile.  The 
horizontal dashed line indicates the level of background
light intensity for each object.
\label{tidaltails}}
\end{figure}

\clearpage
\begin{figure}
\plotone{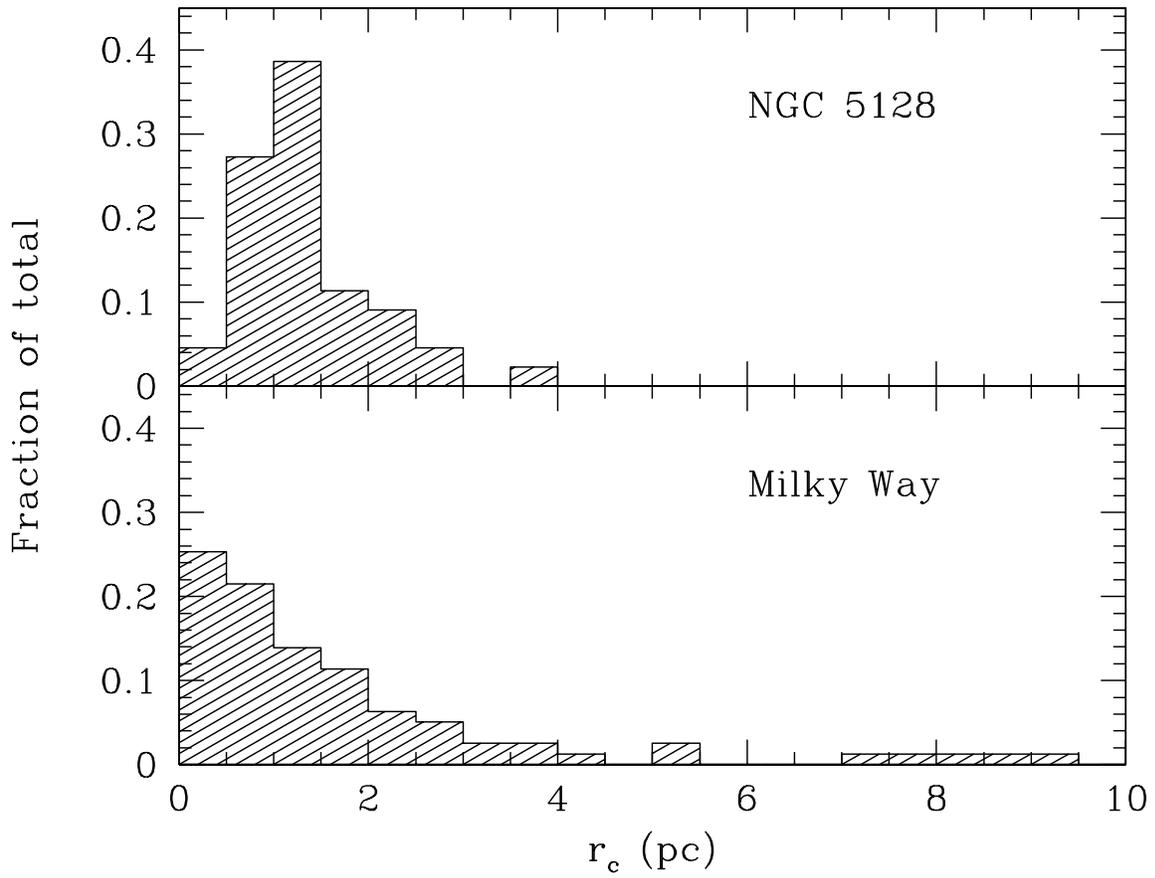}
\caption{Distribution of core radii $r_c$ of NGC 5128 clusters,
compared with those in the Milky Way (see text for definition of
samples).  Our NGC 5128 sample lacks clusters with very small 
core radii, which we believe to indicate selection and measurement bias.
The lack of clusters at very large $r_c$ ($\gtsim 5$ pc) may also
be due to selection bias (see text).
\label{rc_comp}}
\end{figure}

\clearpage
\begin{figure}
\plotone{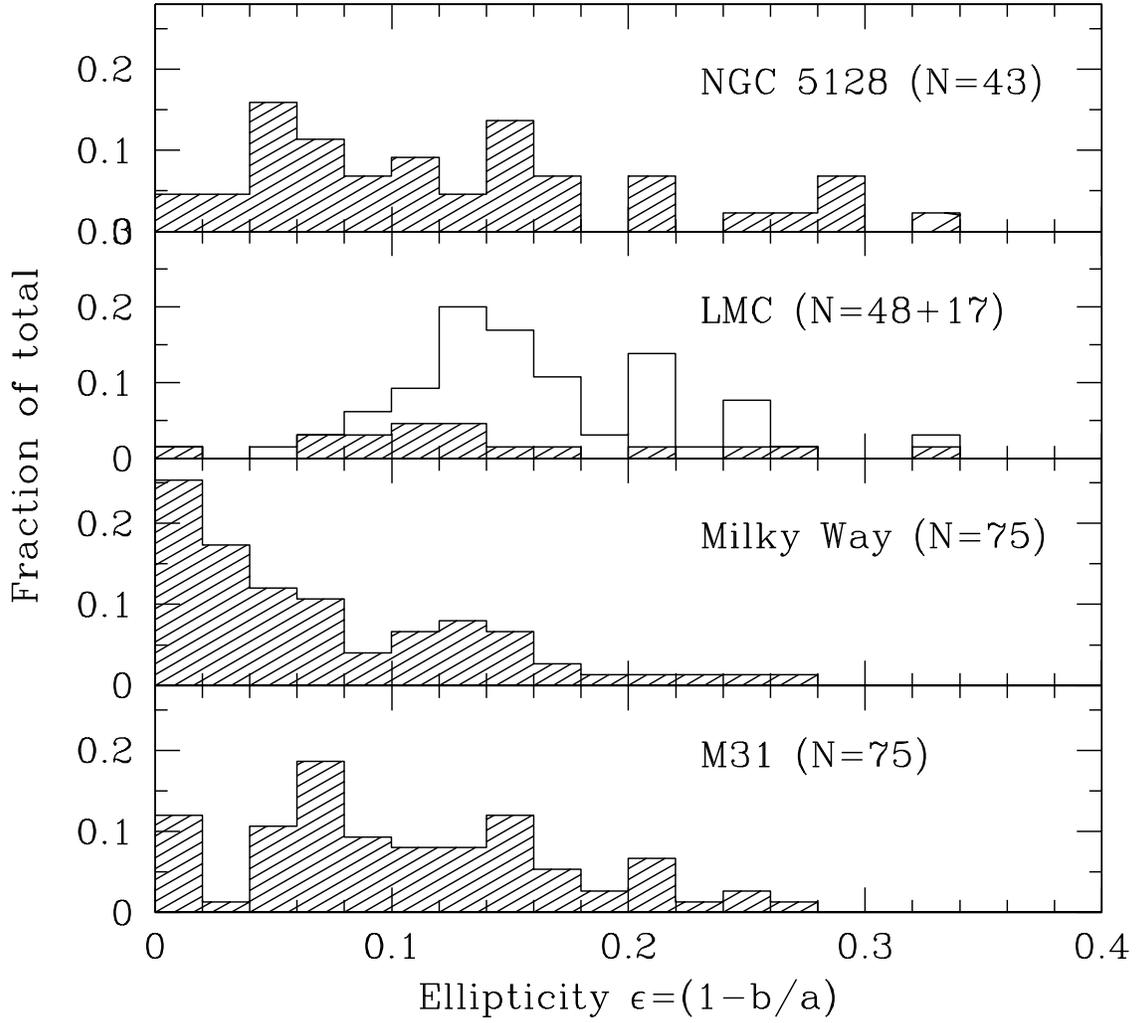}
\caption{Distribution of ellipticities $\epsilon = (1-b/a)$ for clusters in
four galaxies (see text for data sources).  For the LMC, the unshaded
histogram is the distribution for young clusters, whereas
the shaded region is for ``old'' clusters (SWB classes VI-VII).
\label{e_comp}}
\end{figure}

\clearpage
\begin{figure}
\plotone{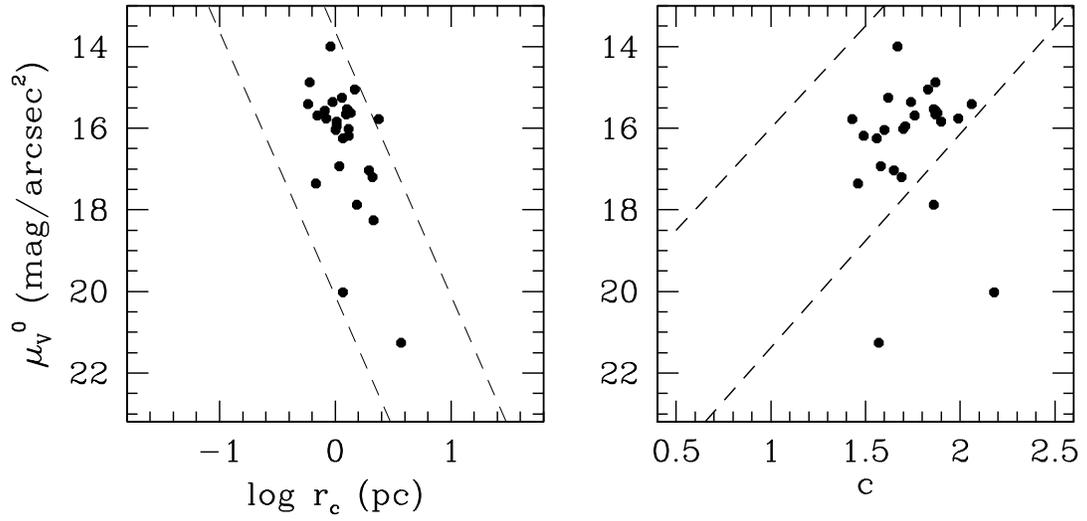}
\caption{Central surface brightness $\mu_V^0$ for our measured clusters,
plotted against core radius (left panel) and central concentration
(right panel).  The dashed lines indicate the regions occupied by
the globular clusters in the Milky Way and M31.
\label{mu_v0}}
\end{figure}

\clearpage
\begin{figure}
\plotone{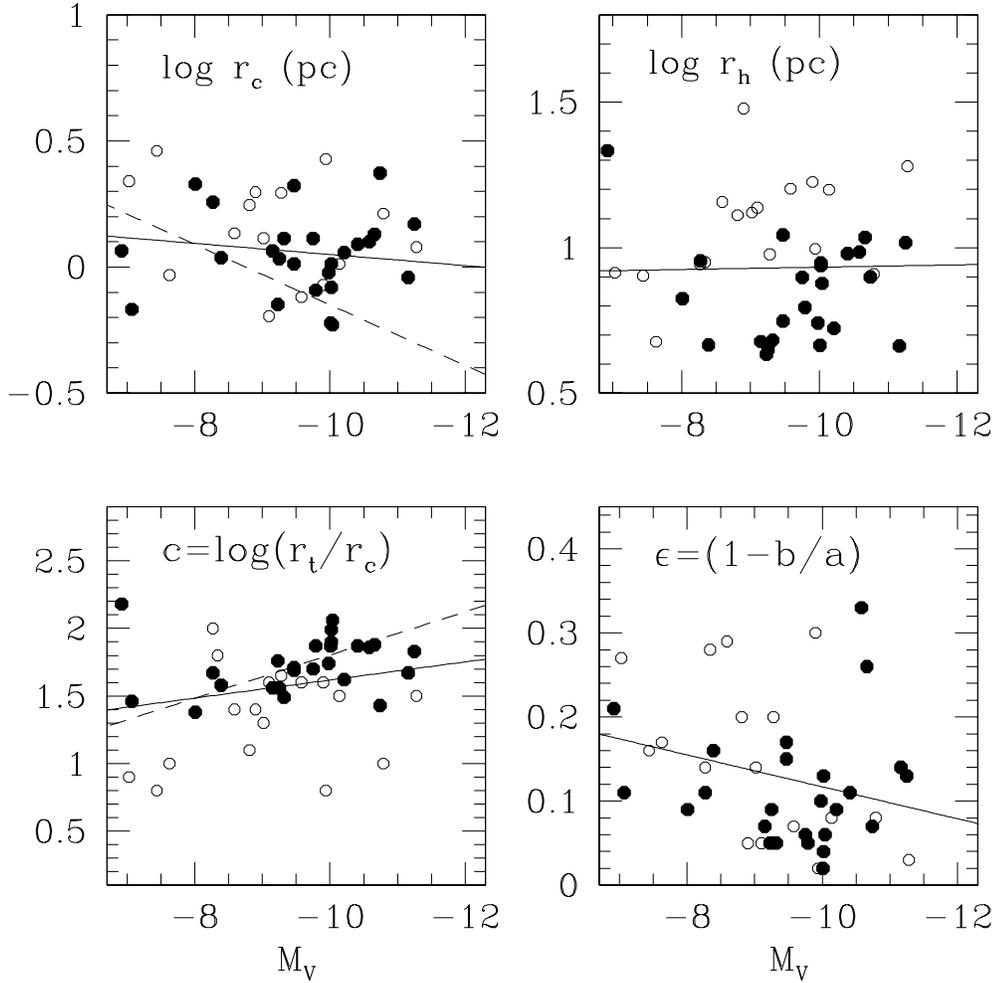}
\caption{Measured structural parameters plotted against
cluster luminosity.  {\sl Upper left:} logarithm of core radius (parsecs).
{\sl Upper right:} logarithm of half-light radius (parsecs).
{\sl Lower left:} central concentration $c$.
{\sl Lower right:} cluster ellipticity $\epsilon$.
In all the graphs, {\sl filled circles} represent measurements from
the higher-resolution STIS or PC1 cameras, while the {\sl open circles}
represent measurements from the lower-resolution WF2,3,4 cameras.
The dashed lines in the two left panels are the mean relations
for the Milky Way clusters, from McLaughlin (2000).
\label{mv_corr}}
\end{figure}

\clearpage
\begin{figure}
\plotone{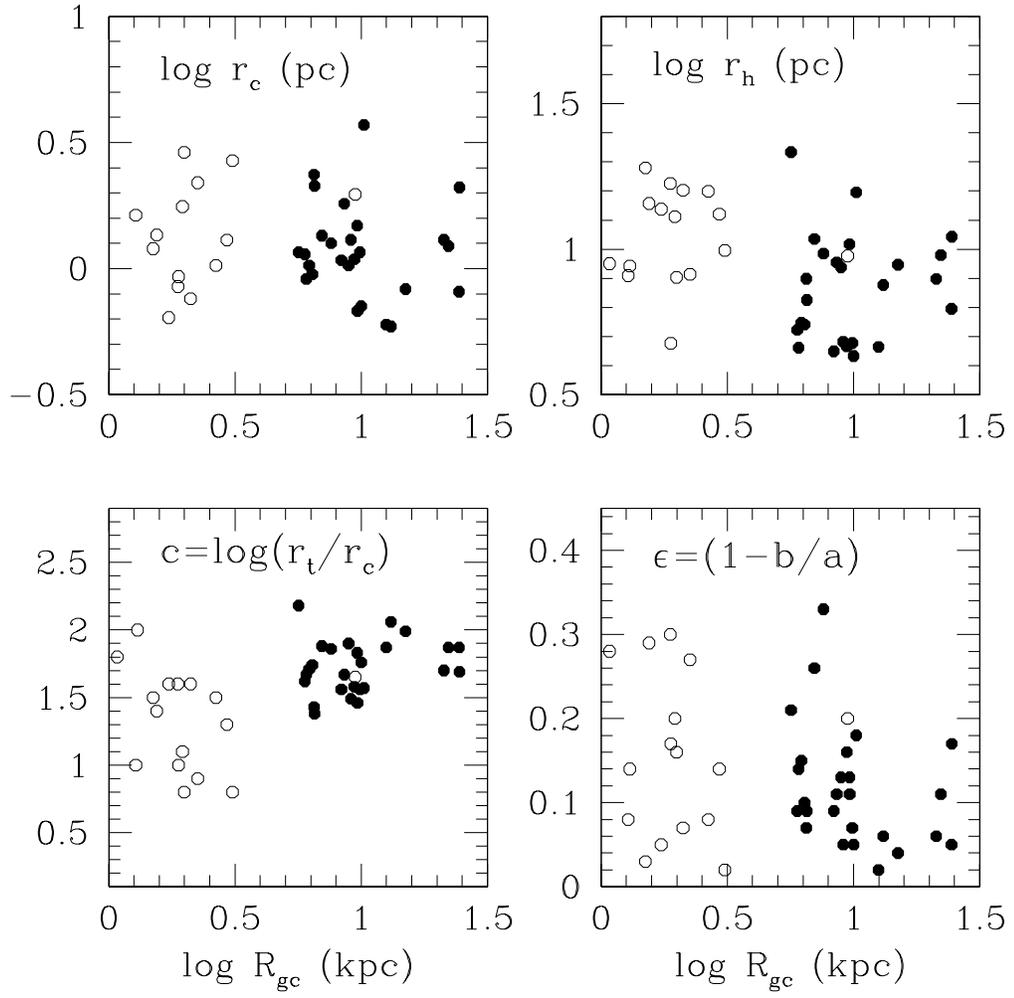}
\caption{Measured structural parameters plotted against 
projected galactocentric distance.  Symbols are as in the
previous figure.
\label{rgc_corr}}
\end{figure}

\clearpage
\begin{figure}
\plotone{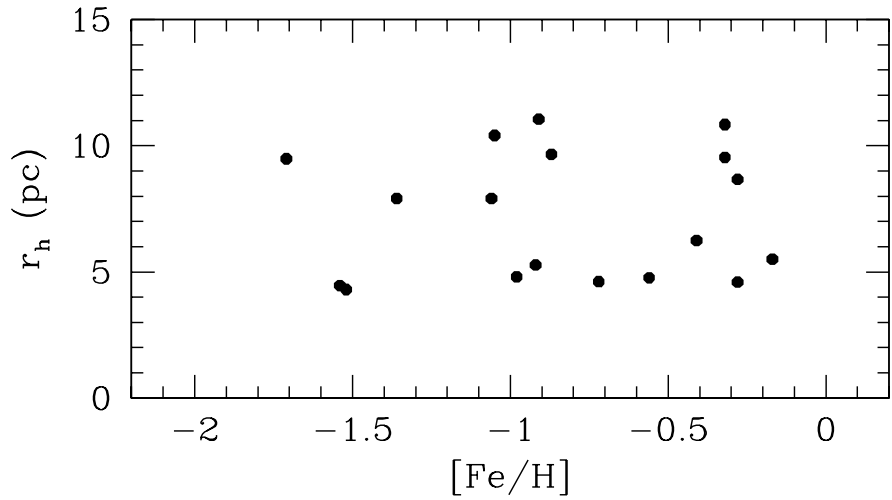}
\caption{Half-mass radius $r_h$ plotted
against cluster metallicity as measured from the
Washington $(C-T_1)$ photometry of Harris et al.~ (1992).
No significant correlation is seen.  Note that $r_h$ here
is expressed in parsecs, whereas in the text the least-squares
fitting equation refers to $r_h$ in arcseconds.  The conversion
factor used is $1'' = 19.39$ pc.
\label{feh_rh}}
\end{figure}

\clearpage
\begin{figure}
\plotone{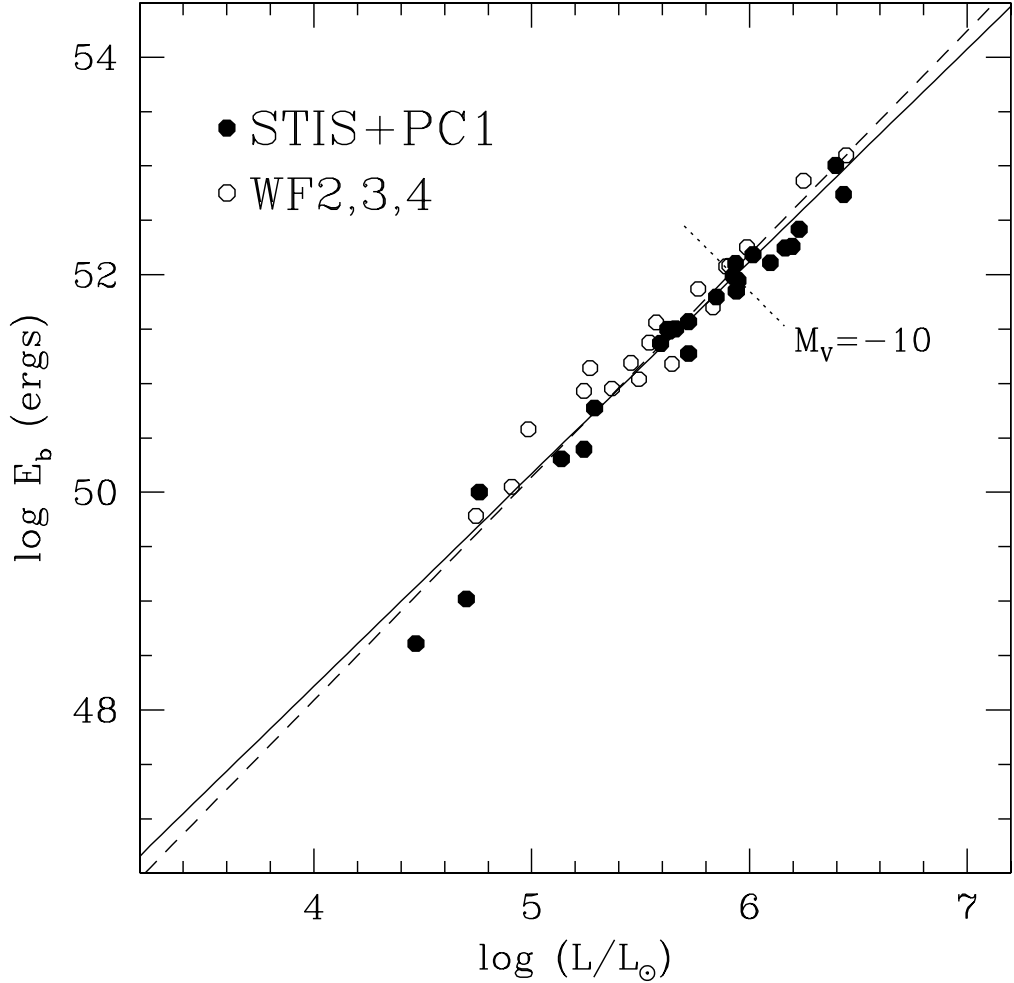}
\caption{Globular clusters in NGC 5128 plotted on the 
plane of binding energy versus cluster luminosity.
Solid dots denote clusters measured from the high-resolution
STIS or PC1 detectors, while open circles denote ones measured
with the lower-resolution WF detectors on the WFPC2 camera.
The {\sl solid line} is the least-squares correlation of the 43
data points shown, while the {\sl dashed line} is the correlation
obtained by McLaughlin (2000a) for the Milky Way clusters.  
The short dotted line marked $M_V = -10$ indicates the rough
upper limit for the brightest Milky Way clusters.
\label{ebplane}}
\end{figure}

\clearpage
\begin{figure}
\plotone{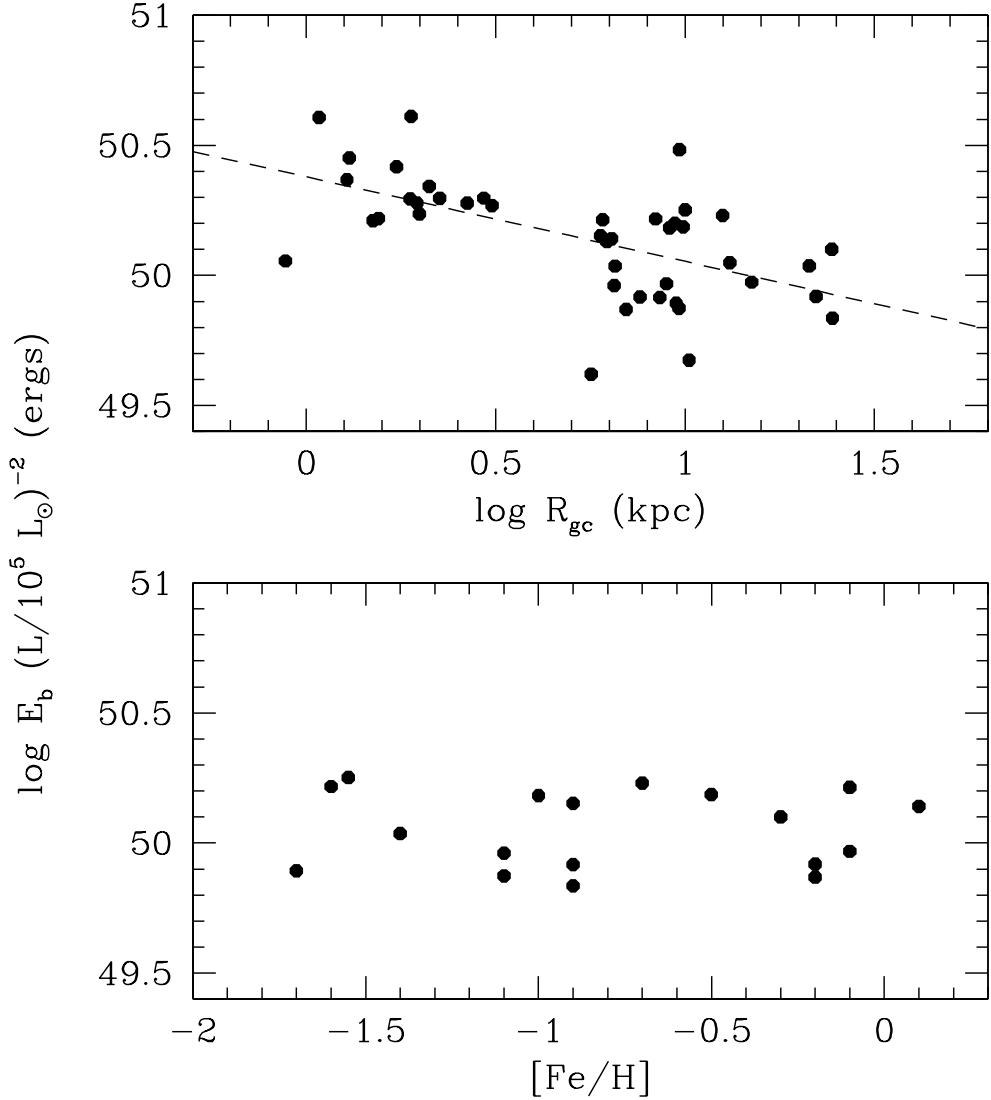}
\caption{{\sl Upper panel:} Binding energy ratio $E_b / L^2$, plotted against
projected galactocentric distance.  The dashed line gives the residual
correlation $(E_b/L^2) \sim R_{gc}^{0.325}$ discussed in the text.
{\sl Lower panel:} Binding energy ratio plotted against cluster metallicity.
No residual correlation is seen.
\label{eb_resid}}
\end{figure}

\clearpage
\begin{figure}
\plotone{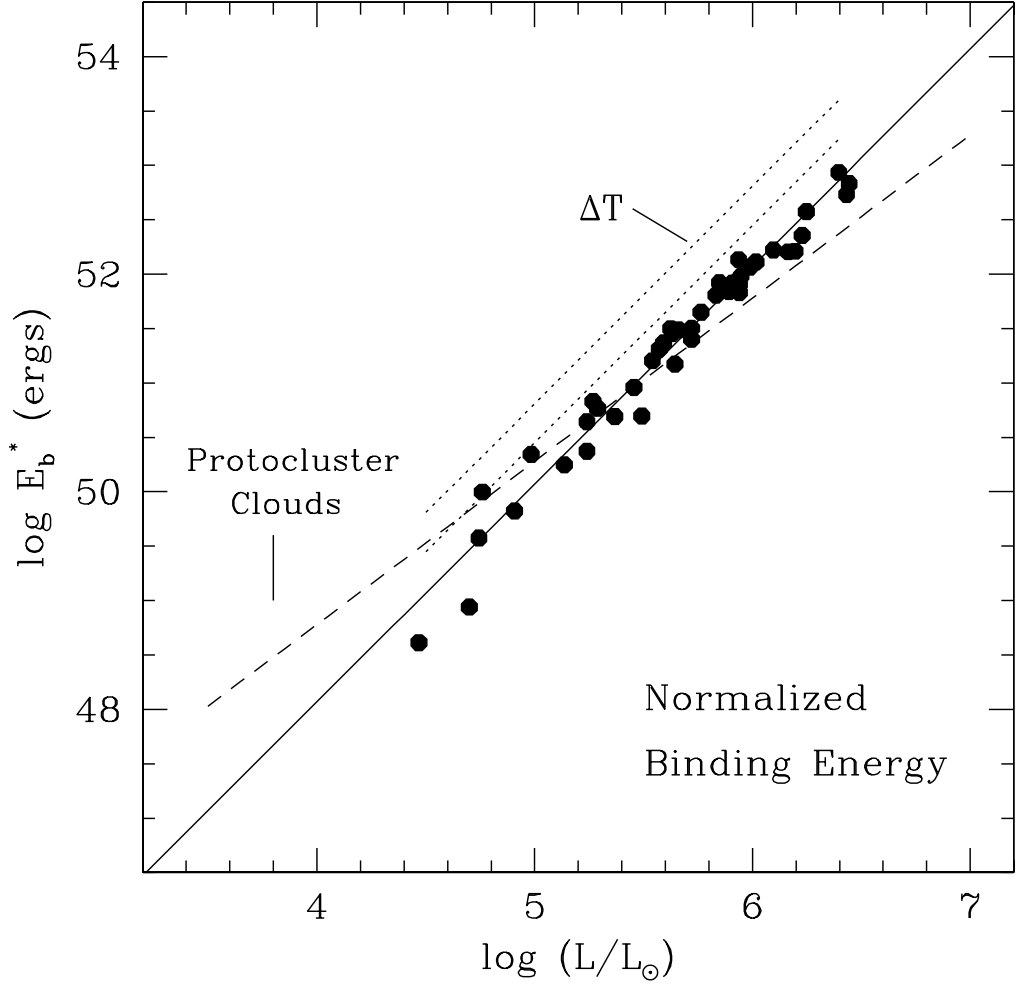}
\caption{Normalized binding energy $E_b^{\star} = 
E_b \cdot (R_{gc}/10 {\rm kpc})^{0.325}$ plotted versus luminosity.
The solid line gives the least-squares relation 
log $E_b^{\star} = 40.07 + 2.00$ log $(L/L_{\odot})$.
The two short dotted lines above it are the loci
for clusters of ages $T=2$ Gyr (upper line) or 5 Gyr (middle line).  
The long dashed line is the equivalent $E_b$ for protocluster
gas clouds in an isothermal potential well at $R_{gc} = 10$ kpc and
for a circular velocity $V_c = 245$ km s$^{-1}$ (see text).
\label{ebstar}}
\end{figure}

\clearpage
\begin{figure}
\plotone{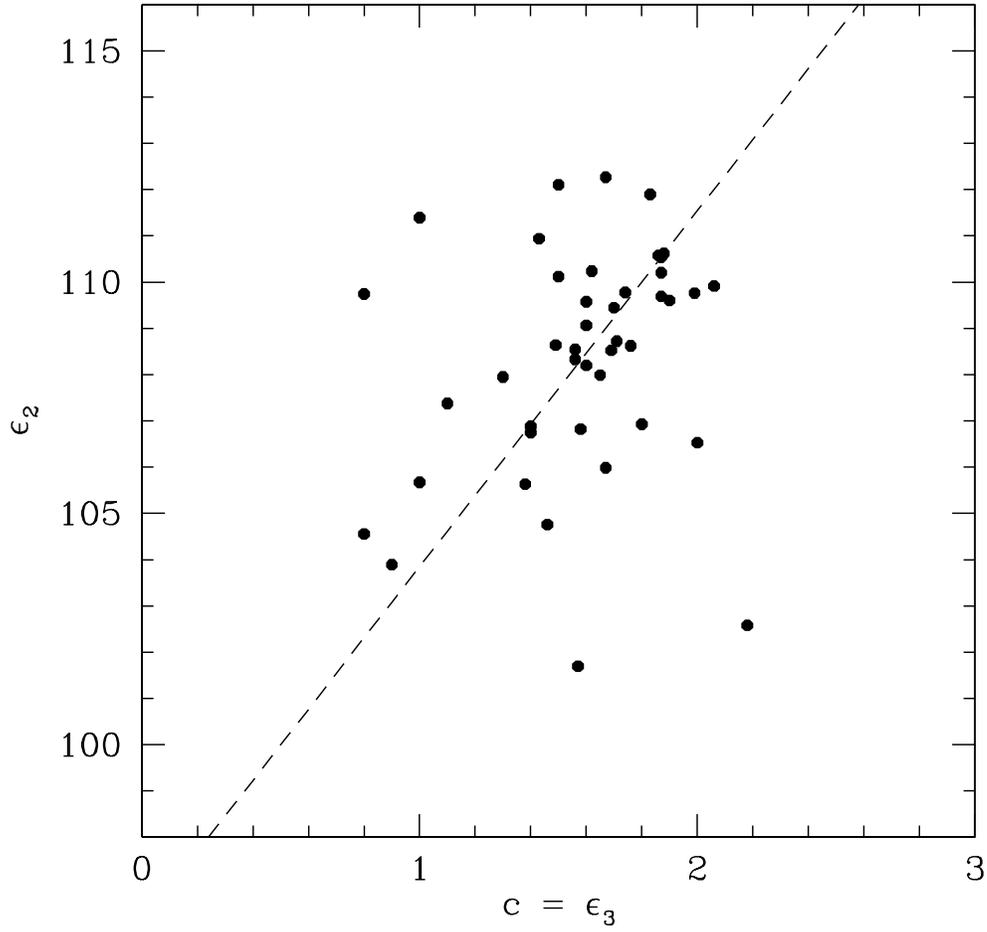}
\caption{A face-on view of the ``fundamental plane'' for
globular clusters.  Here $\epsilon_2 = $ log$(E_b^{\star 2} \cdot L)$
as defined in the text, and $\epsilon_3 = c$.  The 
dashed line gives McLaughlin's (2000a) relation for the Milky Way
clusters, which extends to lower luminosity than we have observed here.
\label{fp}}
\end{figure}

\clearpage
\end{document}